\documentclass[11pt]{article}
\usepackage[normalem]{ulem}
\pdfoutput=1 

\usepackage{jheppub} 
\usepackage{url}
\usepackage{caption}
\usepackage{subcaption}
\usepackage{extarrows}

\usepackage[latin9]{inputenc}
\usepackage{float}
\usepackage{amsmath}
\usepackage{amssymb}
\usepackage{graphicx}
\usepackage{esint}
\usepackage{hyperref}
\usepackage{comment}
\usepackage{color}
\usepackage{microtype}
\usepackage{cleveref}
\usepackage{breakurl}
\usepackage{bbm}
\newcommand{\be}{\begin{equation}}
\newcommand{\ee}{\end{equation}}
\newcommand{\ben}{\begin{displaymath}}
\newcommand{\een}{\end{displaymath}}
\newcommand{\bea}{\begin{eqnarray}}
\newcommand{\eea}{\end{eqnarray}}
\def\K{K{\"a}hler }
   \newcommand{\rf}[1]{(\ref{#1})}
\newcommand{\vp}{\varphi}

\def\be{\begin{equation}}
\def\ee{\end{equation}}
\def\bea{\begin{eqnarray}}
\def\eea{\end{eqnarray}}
\def\ba{\begin{array}}
\def\ea{\end{array}}
\def\bit{\begin{itemize}}
\def\eit{\end{itemize}}

\def\a{\alpha}

\def\vp{\varphi}

 \makeatletter

\allowdisplaybreaks

\makeatother

\makeatletter
\DeclareRobustCommand{\rcite}[1]{%
  \rcite@aux#1,\@nil{#1}%
}
\def\rcite@aux#1,#2\@nil#3{%
  \if\relax#2\relax
    Ref.~\cite{#3}%
  \else
    Refs.~\cite{#3}%
  \fi
}
\makeatother

\hypersetup{
    colorlinks = true,
    citecolor = {blue},
    linkcolor = {blue},
    urlcolor = {blue},
}

 \title{\rm { \LARGE \bf   On the Present Status of Inflationary Cosmology
}}

\author{Renata Kallosh and Andrei Linde}

\affiliation{Leinweber Institute for Theoretical Physics at Stanford, 382 Via Pueblo, Stanford, CA 94305, USA}
\emailAdd{kallosh@stanford.edu}
\emailAdd{alinde@stanford.edu}

\abstract{We give a brief review of the basic principles of inflationary theory and discuss the present status of the simplest inflationary models which can describe Planck/BICEP/Keck observational data by choice of a single model parameter.  In particular, we discuss the Starobinsky model, Higgs inflation,  and $\alpha$-attractors, including the recently developed $\alpha$-attractor models with $SL(2,\mathbb{Z})$ invariant potentials. We also describe inflationary models providing a good fit to the recent ACT data, as well as the polynomial chaotic inflation models with three parameters, which can account for any values of the three main CMB-related inflationary parameters $A_{s}$, $n_{s}$ and $r$.}

\begin{document}

\maketitle

 
\parskip 7 pt

\section{Introduction}
The inflationary theory was invented at the beginning of the 80s to address numerous problems of the standard Big Bang cosmology  \cite{Starobinsky:1980te, Guth:1980zm,Linde:1981mu,Linde:1983gd,Linde:1986fd}. Over the last 40 years, this theory has evolved from a collection of bold scientific proposals into the cosmological paradigm that describes the origin of the universe and the formation of its large-scale structure.

The first realistic model of inflationary type was proposed by Starobinsky \cite{Starobinsky:1980te}. However, it did not attempt to solve the homogeneity, isotropy, horizon, and flatness problems, which were the main goals of the inflationary scenario proposed by Guth  \cite{Guth:1980zm}. Therefore, initially, the Starobinsky model remained slightly outside the mainstream of inflationary cosmology; however, it was later realized that the predictions of this model, derived by Mukhanov and Chibisov \cite{Mukhanov:1981xt}, provided a perfect match to the WMAP and Planck data.  

The scenario proposed by Guth beautifully formulated the main goals of inflationary cosmology, but had a problem which was well recognized by its author \cite{Guth:1980zm}: the universe became grossly inhomogeneous after the end of inflation. For a while, this problem seemed to be unsolvable \cite{Hawking:1982ga, Guth:1982pn}.  A possible solution was proposed in the new inflationary scenario \cite{Linde:1981mu}, but it had its own share of problems.  

Old and new inflation \cite{Guth:1980zm,Linde:1981mu} represented only a partial modification of the hot Big Bang theory. It was still assumed that the
universe was in a state of thermal equilibrium from the very
beginning, that it was relatively homogeneous and large enough
even before inflation, and that the stage of
inflation was just an intermediate stage of the evolution of the
universe. Long ago, these textbook assumptions seemed
practically unavoidable.  That is why it was difficult to overcome a 
psychological barrier and abandon all of these assumptions. This
was done with the invention of the chaotic inflation scenario
\cite{Linde:1983gd}.  According to this scenario, inflation may begin even if there was no thermal
equilibrium in the early universe, and it may occur even in
the theories with the simplest potentials such as $V(\phi) \sim
\phi^2$. Moreover, this scenario is not limited to the theories with monomial or polynomial potentials: chaotic inflation occurs in {\it any} theory where the potential has a sufficiently flat region, which allows the existence of the slow-roll regime \cite{Linde:1983gd}.

In this paper, we will describe some of the basic and most popular inflationary models, discuss their cosmological predictions, and their status in view of the latest observational data.

\section{Chaotic inflation}
We begin with the simplest inflationary model describing a single massive scalar field with the Lagrangian
 \be
{ {\cal L} \over \sqrt{-g}} =  {R\over 2}  -  {(\partial_{\mu} \phi)^2\over 2} - {m^{2}\over 2} \phi^{2}  \,  .
\label{chaot}
\ee
The first term is the gravitational Lagrangian, the second is the kinetic term of the scalar field $\phi$, and the third term describes the potential of a scalar field with mass $m$ \cite{Linde:1981mu}. It was a great surprise that this simple theory could describe inflation.

Since this function has a minimum at $\phi = 0$,  one could
expect that the scalar field $\phi$ should oscillate near this
minimum. This is indeed the case if the universe does not expand.
 However, because of the expansion of the universe with Hubble
constant $H = \dot a/a $, an additional  term $3H\dot\phi$ appears in the harmonic oscillator equation:
\begin{equation}\label{1x}
 \ddot\phi + 3H\dot\phi = -m^2\phi \ .
\end{equation}
This equation looks like an equation for the harmonic oscillator with the friction term $3H\dot\phi$. 

The Einstein equation for a homogeneous universe containing a scalar
field $\phi$ is 
\begin{equation}\label{2x}
H^2 +{k\over a^2} ={1\over 6}\, \left(\dot \phi ^2+m^2 \phi^2)
\right) \ .
\end{equation}
Here $k = -1, 0, 1$ for an open, flat or closed universe
respectively. We work in units $M_p^{-2} = 8\pi G = 1$.

If   the scalar field $\phi$  initially was large,   the Hubble
parameter $H$ was large, the friction term $3H\dot\phi$ was large, and, therefore, the scalar field was moving very slowly. At this stage, the energy density of the scalar field, unlike the  density of ordinary matter, remained
almost constant, and the expansion of the universe continued with a greater speed than in the old Big Bang theory. Due to the
rapid growth of the scale of the universe and the slow motion of the
field $\phi$, soon after the beginning of this regime, one has
$\ddot\phi \ll 3H\dot\phi$, $H^2 \gg {k\over a^2}$, $ \dot \phi
^2\ll m^2\phi^2$, so  the system of equations can be simplified:
\begin{equation}\label{E04}
H= {\dot a \over a}   ={ m\phi\over \sqrt 6}\ , ~~~~~~  \dot\phi =
-m\  \sqrt{2\over 3}     .
\end{equation}
The first equation shows that if the field $\phi$ changes slowly, the size of the universe in this regime grows approximately as
$e^{Ht}$, where $H = {m\phi\over\sqrt 6}$. This is the stage of
inflation, which ends when the field $\phi$ becomes much smaller
than $O(1)$. The solution of these equations shows that after a long
stage of inflation, the universe of initial size $a_{0}$  filled with the field
$\phi_{0}   \gg 1$  grows  exponentially \cite{Linde:1990flp} and becomes given by $a_{e}=a_0 e^{N_{e}} $, where  
\begin{equation}\label{E04aa}
a_{e} =  a_0 e^{N_{e}} =   e^{\phi_{0}^2/4} \ .
\end{equation}
In particular, the universe expands by more than $e^{60}$ times during inflation if
\be\label{mag}
\phi_{0} > 15.5 \ .
\ee

Thus, inflation does not require an initial state of thermal equilibrium, 
supercooling, and tunneling from the false vacuum, which seemed to be a necessary requirement of some of the first versions of inflationary cosmology \cite{Guth:1980zm,Linde:1981mu}. Inflation appears to be a generic cosmological regime, which is possible even in theories that can be as simple as a theory of a harmonic oscillator \cite{Linde:1983gd}. A similar inflationary regime is possible in any theory where the potential is not too steep. This includes {\it all} \ models to be discussed in our paper.

\section{Initial conditions for inflation}\label{ini}

The primary issue with the old Big Bang theory was that cosmological evolution was supposed to be approximately adiabatic, and the total number of particles in the universe was supposed to be approximately conserved, changing by no more than one or two orders of magnitude. From this fact, one could show that at the Planckian time $t = O(1)$  when the density of the hot universe was Planckian $\rho = O(1)$, it consisted of about $10^{{90}}$ causally disconnected parts of a Planckian size $l = O(1)$, each of which contained only $O(1)$ elementary particles. In this situation, large-scale homogeneity and isotropy of the universe would appear miraculous, and the origin of the exponentially large number of particles of the universe would seem equally mysterious.

To explain how inflationary theory solves this problem, let us consider a closed universe of the smallest initial size $l = O(1)$, which emerges in a state
with the Planck density $\rho = O(1)$. So far, this is very similar to the assumptions made in the Big Bang theory. The major difference is that now, instead of considering $10^{{90}}$  parts of the universe with such properties, it is sufficient to consider one such tiny Planckian-size domain.

The condition that the universe emerged with the Planck density $\rho = O(1)$ implies that at this initial moment, the sum of the kinetic energy density, gradient energy
density and the potential energy density is of the order unity:\, ${1\over 2} \dot\phi^2 + {1\over 2} (\partial_i\phi)^2 +V(\phi) \sim 1$.

There are no {\it a priori} constraints on
the initial value of the scalar field in this domain, except for the
constraint ${1\over 2} \dot\phi^2 + {1\over 2} (\partial_i\phi)^2 +V(\phi) \sim
1$.  Consider, for a moment, a theory with $V(\phi) = const$. This theory
is invariant under the {\it shift symmetry}  $\phi\to \phi + c$. Therefore, in such a
theory {\it all} initial values of the homogeneous component of the scalar field
$\phi$ are equally probable.  

The only constraint on the initial value of the field appears if the effective potential is not constant but grows and becomes greater
than the Planck density at $\phi > \phi_p$, where  $V(\phi_p) = 1$. This
constraint implies that $\phi \lesssim \phi_p$,
but there is no model-independent reason to expect that initially
$\phi$ must be much smaller than $\phi_p$. In the context of the realistic version of the model with the potential $ {m^{2}\over 2} \phi^{2} $  with $m = O(10^{{-5}})$ this implies that the upper constraint on the initial value of the field $\phi$ is $\phi \lesssim 1/m \sim 10^{5}$.

Thus, we expect that typical initial conditions correspond to
${1\over 2}
\dot\phi^2 \sim {1\over 2} (\partial_i\phi)^2\sim V(\phi) = O(1)$.
If ${1\over 2} \dot\phi^2 + {1\over 2} (\partial_i\phi)^2
 \lesssim V(\phi)$ in the part of the universe under consideration, one can show that inflation begins,
and then within the Planck time the terms  ${1\over 2} \dot\phi^2$ and ${1\over 2}
(\partial_i\phi)^2$ become much smaller than $V(\phi)$, which ensures
continuation of inflation.  The conclusion is that inflation
occurs under rather natural initial conditions if it can begin at $V(\phi)
= O(1)$ \cite{Linde:1983gd,Linde:1985ub,Linde:2005ht}.  

During the last decade, the problem of initial conditions for inflation was thoroughly investigated by sophisticated numerical methods of the general theory of relativity \cite{Carrasco:2015rva,East:2015ggf,Kleban:2016sqm,Clough:2016ymm,Linde:2017pwt,Clough:2017efm,Aurrekoetxea:2019fhr,Creminelli:2020zvc,Joana:2020rxm,Joana:2022pzo,Corman:2022alv,Elley:2024alx,Aurrekoetxea:2024mdy,Joana:2024ltg}. All of these papers not only confirmed the validity of the basic principles of chaotic inflation, but also found that inflation can naturally begin even in the models with $V(\phi) \ll 1$.\footnote{A notable exception is Ref.  \cite{Garfinkle:2023vzf} claiming that if the initial value of potential energy density of the inflaton ${m^{2}\over 2}\phi^{2}$ is many orders of magnitude smaller than $O(1)$ and its kinetic and gradient energy density and the Weyl tensor are $O(1)$, inflation does not last sufficiently long, $N_{e} < 60$. It was argued in \cite{Garfinkle:2023vzf}  that all other authors \cite{Carrasco:2015rva,East:2015ggf,Kleban:2016sqm,Clough:2016ymm,Linde:2017pwt,Clough:2017efm,Aurrekoetxea:2019fhr,Creminelli:2020zvc,Joana:2020rxm,Joana:2022pzo,Corman:2022alv,Elley:2024alx,Aurrekoetxea:2024mdy,Joana:2024ltg} reached an opposite conclusion because they did not consider sufficiently general initial conditions. However, the real reason was much simpler: The initial value of $\phi$ considered in \cite{Garfinkle:2023vzf} did not satisfy the basic requirement \rf{mag}. In their next paper \cite{Ijjas:2024oqn}, the authors increased $\phi_{0}$ and no longer claimed that $N_{e}< 60$.}


The simplest possibility to describe inflation at $V \ll  1 $ and/or $\phi \ll 1$ is to consider multi-field inflationary models.  For example, one may consider a model 
\be\label{simpleinit}
V(\phi, \chi) = {m^{2}\over 2} \chi^{2}  +V(\phi) \ ,
\ee
where $V(\phi)$ is a plateau potential of the type which appears in the $\alpha$-attractor models to be considered in this paper.  Inflation driven by the field $\chi$ may begin at  ${m^{2}\over 2} \chi^{2} = O(1)$, as in the simplest chaotic inflation model discussed above. If the field $\phi$ initially was at the plateau of its potential $V(\phi)$, it did not move until the end of the first stage of inflation driven by the field $\chi$. After that, the stage of inflation driven by the field $\phi$ begins, and if this stage is long enough, it determines the formation of the large-scale structure and CMB perturbations in the observable part of the universe. Various versions of this scenario have been discussed in \cite{Linde:1987yb,Linde:2014nna,Linde:2017pwt,Corman:2022alv}. We will revisit this topic in this paper, within the context of the Starobinsky model.

\section{Can we test inflation?}
The best way to test inflation is to study the properties of cosmic microwave background radiation (CMB).
Here is a list of the most important predictions of inflationary cosmology confirmed by observations.
\begin{enumerate}

\item The universe is nearly exactly homogeneous and isotropic. 45 years ago, when the inflationary theory was first developed, its homogeneity was known only approximately. Since that time, homogeneity and isotropy of the universe have been established with a much greater precision.   

\item The universe must be almost exactly flat, $\Omega = 1$. Most inflationary models predict $\Omega = 1 \pm 10^{{-4}}$. 45 years ago, when we
only knew that  $\Omega = O(1)$.  In the mid-1990s, the consensus was that $\Omega=0.3$, until the discovery of dark energy, which allowed $\Omega = 1$, consistent with inflation.   The latest observational results show that $\Omega = 1 \pm 10^{{-2}}$, in agreement with
inflationary predictions. 

\item In the early 80's, some authors argued that inflation was ruled out because adiabatic perturbations are not observed at the expected level $O(10^{-3})$ required for galaxy formation. Thanks to dark matter, smaller perturbations are sufficient, and they were found by COBE.

\item Perturbations of the metric produced during inflation are adiabatic.  

\item These perturbations are approximately Gaussian. In non-inflationary models, the parameter  $f_{NL}$ describing the level of the so-called local non-Gaussianity can be as large as $O(10^{4})$, but it is predicted to be $O(1)$ in all single-field inflationary models. Prior to the Planck 2013 data release, rumors circulated that $f_{NL} = O(30)$, which would rule out most single-field inflationary models.
The latest results confirm that $|f_{NL}|  \lesssim 5$.  

\item Inflationary perturbations should have a nearly
flat spectrum with the spectral index $n_{s}$ close to $1$. The latest results show that  $n_{s} \sim 0.96-0.98$.  

\item On the other hand, the spectrum of inflationary perturbations was predicted to be slightly non-flat. A small deviation of $n_{s}$ from 1 is one of the distinguishing features of inflation. It is as significant for inflationary theory as the asymptotic freedom for the theory of strong interactions. Indeed,  the possibility $n_s=1$ is strongly disfavored by the CMB-related data.   

\item Inflation does not produce vector perturbations. Indeed, they have not been found.   

\item Inflationary perturbations should produce specific peaks in the spectrum of CMB radiation. Indeed,
many such peaks have been found.  The existence of such peaks ruled out alternative theories of the universe's large-scale structure in the context of cosmic string and texture theories.

\item The large angle TE anti-correlation (WMAP, Planck) is a distinctive signature of superhorizon fluctuations \cite{Spergel:1997vq}, ruling out many alternative possibilities.  
\end{enumerate}

\begin{figure}[!h]

\begin{center}
\hskip -0.3cm
\includegraphics[width=9.5cm]{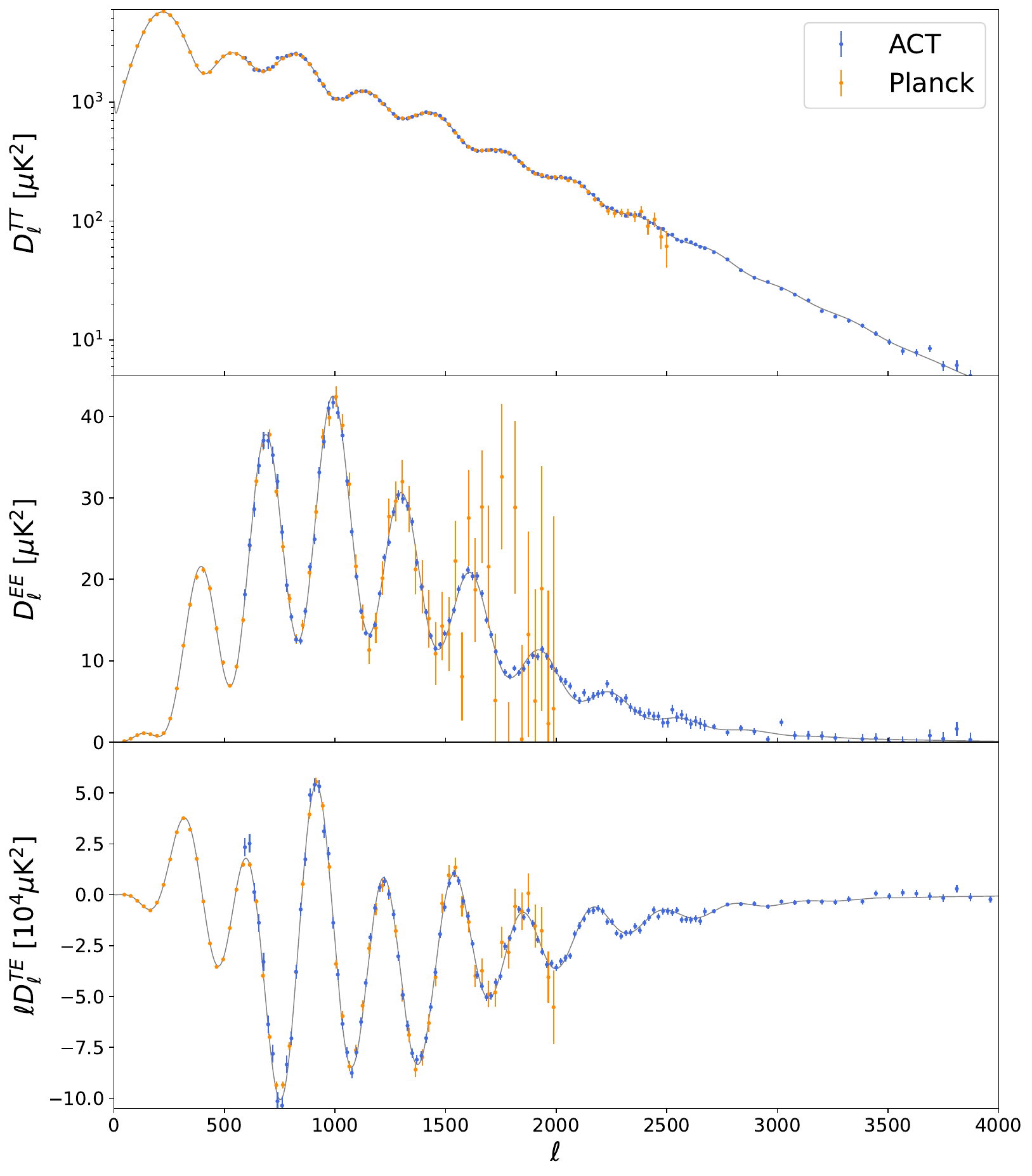}
\caption{\footnotesize ACT DR6 and Planck PR3 (Planck Collaboration 2020b) combined TT (top), EE (middle), and TE (bottom) power
spectra. The gray lines show the joint ACT and Planck ${\rm \Lambda CDM}$ best-fit power spectra.   }
\vskip -15pt
\label{ACTnsr}
\end{center}
\end{figure}

Thus, many basic principles and predictions of inflationary theory have been tested and confirmed.  The present state of the art in this area can be illustrated by the first figure from the recent data release of the Atacama Cosmology Telescope (ACT) \cite{ACT:2025fju}, which is reproduced in Fig. \ref{ACTnsr} of this review. Amazingly, one can describe all the tiny features of the plots shown in Fig. \ref{ACTnsr} in the context of simple inflationary models with one or two parameters. 

The next stage of investigation, discussed below, involves examining the predictions of specific inflationary models and comparing them with one another.

\

\section{Constraints on inflationary models}\label{constr}


One of the best ways to constrain the large set of available inflationary models is to examine their predictions for the spectral index $n_{s}$ describing the slope of the spectrum of scalar perturbations and also the parameter $r$ describing the ratio of the tensor perturbations to the scalar ones, see \cite{Planck:2013jfk} for a detailed definition of these parameters.
\begin{figure}[!h]
\begin{center}
\hskip  -0.3cm
\includegraphics[width=7cm]{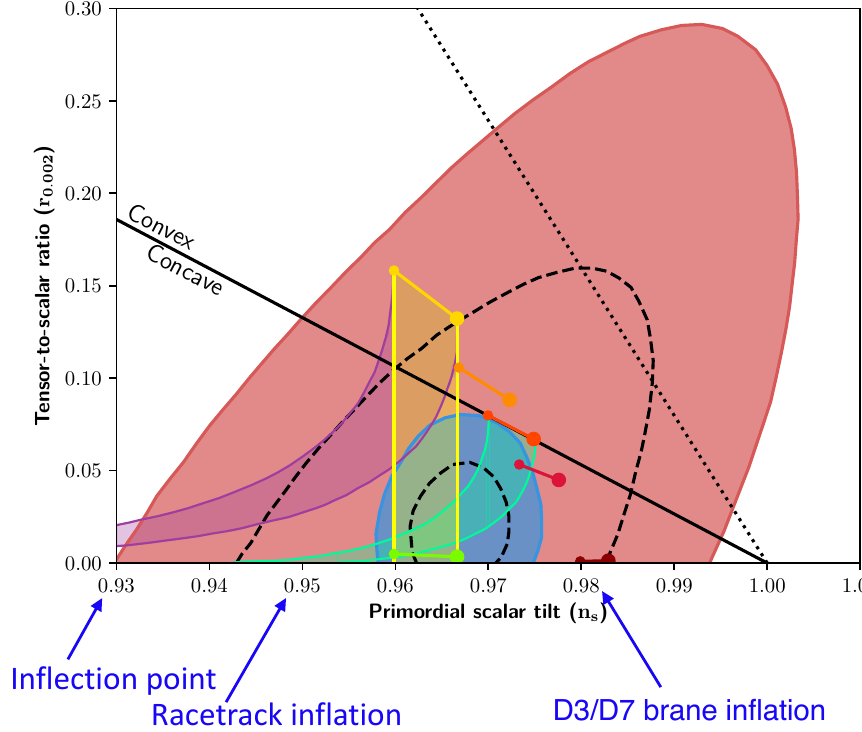}
\caption{\footnotesize  Many favorite string inflation models from a decade ago, such as inflation with an inflection point, Racetrack Inflation, and D3/D7 inflation, were compatible with WMAP (red area) but have been ruled out by Planck 2018 (blue area). The yellow area corresponds to $\alpha$-attractors \cite{Kallosh:2013hoa,Ferrara:2013rsa,Kallosh:2013yoa,Galante:2014ifa,Kallosh:2015zsa,Kallosh:2019eeu,Kallosh:2019hzo}, the two dots in the lower part of this area correspond to the Starobinsky model  \cite{Starobinsky:1980te} and the Higgs inflation model  
  \cite{Salopek:1988qh,Bezrukov:2007ep}. }
\vskip -15pt
\label{nsr}
\end{center}
\end{figure} 
The rapid progress in this direction can be illustrated by Fig. \ref{nsr}. It shows the broad range of allowed values of $n_{s}$ and $r$ in the 7-year WMAP data release, superimposed with the constraints in the Planck 2018 data release.  As one can see, many inflationary models compatible with WMAP have been ruled out by the Planck 2018 results.
 \begin{figure}[h!]
\vskip  5pt
\centering
\includegraphics[scale=0.7]{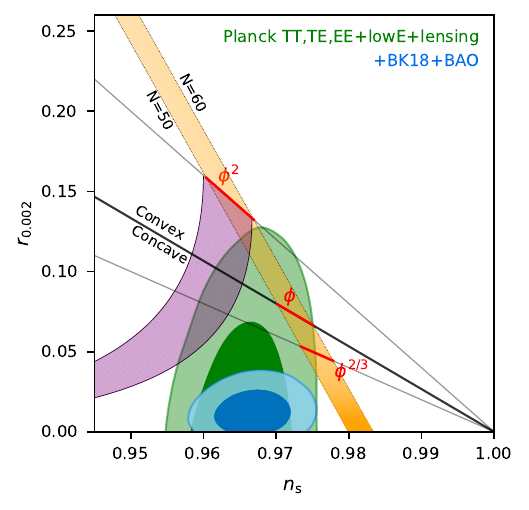}
\vskip -15pt
\caption{\footnotesize  BICEP/Keck results for $n_{s}$ and $r$ \cite{BICEP:2021xfz}. The $1\sigma$  and $2\sigma$  areas are represented by dark blue and light blue colors. The purple region shows natural inflation, and the orange band corresponds to inflation driven by the scalar field with canonical kinetic terms and monomial potentials.}
\label{BICEPKeck}
\end{figure}

The subsequent progress in this direction was quite rapid, and sometimes quite unexpected. Fig. \ref{BICEPKeck} shows that many models allowed by Planck 2018 (green area) have been ruled out by a combination of Planck data and the results obtained by BICEP/Keck \cite{BICEP:2021xfz}.
 In particular, the standard version of natural inflation \cite{Freese:1990rb}, as well as the full class of monomial potentials $V \sim \phi^{n} $, are now strongly disfavored.

We should also discuss here the latest data release of the Atacama Cosmology Telescope (ACT)  \cite{Louis:2025tst,ACT:2025tim}. As we already mentioned, their results provide strong support for inflationary cosmology, see Fig. \ref{ACTnsr}. However, these results,  combined with the results of Planck and of  DESI Y1 data release \cite{DESI:2024uvr,DESI:2024mwx}, significantly modified the constraints on the spectral index $n_{s}$. 

 The constraint given by Planck 2018 was  $n_{s } = 0.9651 \pm  0.0044$ \cite{Planck:2018vyg}. The constraint on $n_{s}$ based on the ACT data is $ n_{s} = 0.9666 \pm  0.0077$, which is perfectly consistent with the Planck 2018 result. However, the new constraint based on a combination of the Planck and ACT data (P-ACT) is $n_{s} = 0.9709 \pm  0.0038$. Finally, a combination of Planck, ACT, and DESI (P-ACT-LB) gives $n_{s} =0.9743 \pm  0.0034 $  \cite{Louis:2025tst}. This last result differs from the original Planck result by about $2\sigma$.
 
  \begin{figure}[H]
\centering
\includegraphics[scale=0.2]{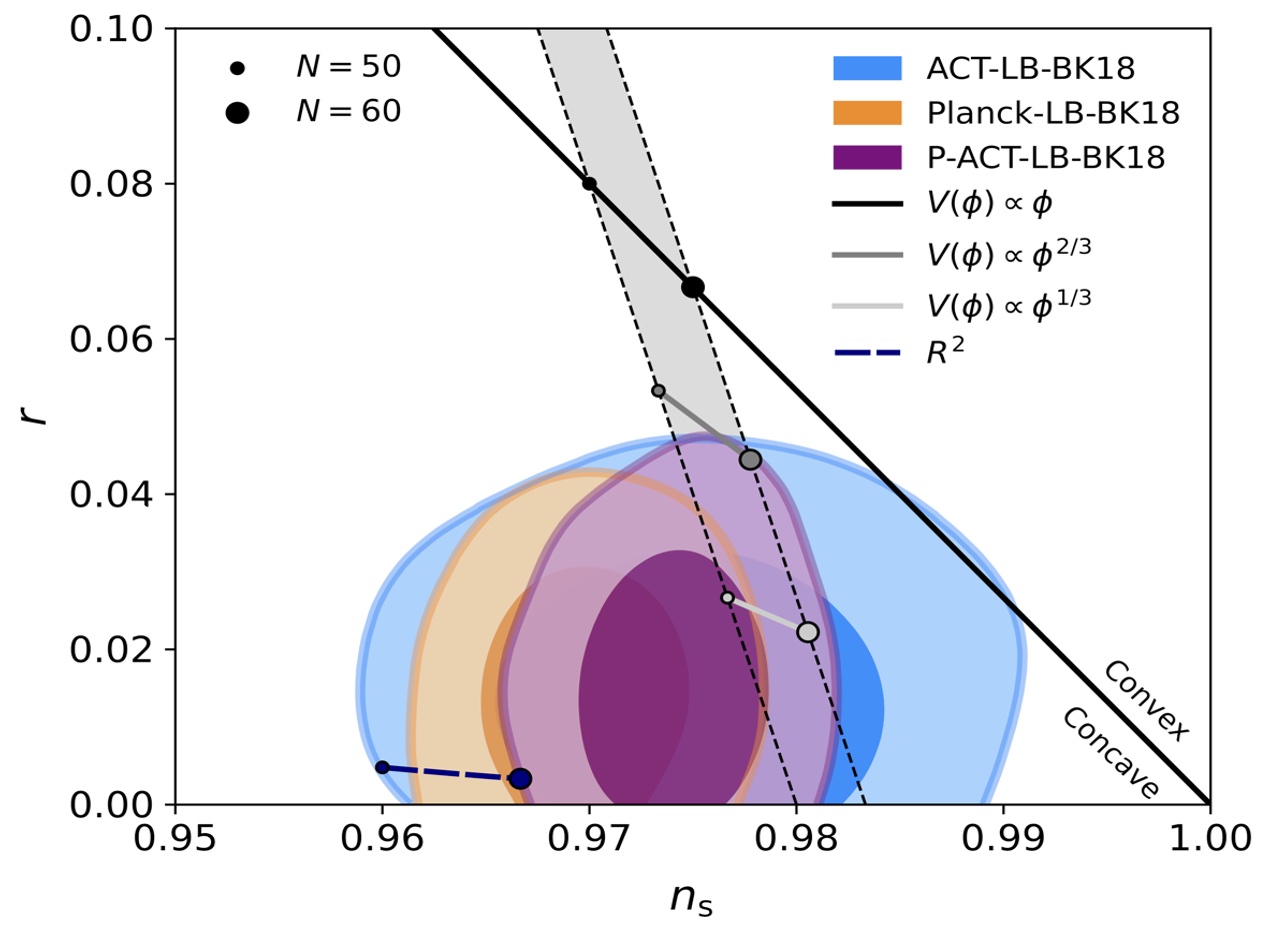}
\vskip -15pt
\caption{\footnotesize  The figure shows the latest constraints on $n_{s}$ and $r$ according to ACT  (P-ACT-LB)  \cite{Louis:2025tst}. The dashed line at the bottom corresponds to the Starobinsky model.}
\label{ACT}
\end{figure}

The P-ACT-LB results are illustrated by Fig. 10 in \cite{ACT:2025tim}; we reproduce it here in Fig. \ref{ACT}. The authors of \cite{Louis:2025tst} conclude, in particular, that the P-ACT-LB constraints on  $n_{s}$ disfavor the Starobinsky model \cite{Starobinsky:1980te} at~$\gtrsim 2\sigma$.  This is a strong and unexpected result, significantly different from the previous results.  The latest SPT data, in combination with all other CMB results plus DESI DR2, give $n_s = 0.9728\pm 0.0027$ \cite{SPT-3G:2025bzu}, which is consistent with the ACT-DESI results.  
 
Note that the constraint $n_{s} =0.9743 \pm  0.0034 $  shown in Fig. 10 in \cite{ACT:2025tim} was obtained after taking into account DESI Y1 results, but assuming $\Omega_{m}+\Omega_{\Lambda} = 1$, i.e. in the $\Lambda$CDM context, see   Table 5 in  \cite{Louis:2025tst}. On the other hand, more recent DESI DR2 results \cite{DESI:2025gwf} strongly disfavor $\Lambda$CDM, which was used in the derivation of this constraint.   The authors of \cite{DESI:2025gwf} say: ``For our baseline DESI DR2 BAO+Planck PR4+ACT likelihood combination, the preference for evolving dark energy over a cosmological constant is about $3\sigma$, increasing to over $4\sigma$  with the inclusion of Type Ia supernova data.''  Thus, one may wonder whether one can disfavor anything at the $2\sigma$ level by using databases that are strongly inconsistent with each other.

Indeed, the discussion of these results in the SPT data release  \cite{SPT-3G:2025bzu} was quite cautious. In particular, they noticed that the constraints  $n_s=0.9739\pm 0.0034$ \cite{ACT:2025fju,ACT:2025tim} and  $n_s = 0.9728\pm 0.0027$ \cite{SPT-3G:2025bzu} were obtained in the context of the $\Lambda$CDM theory, but in this context the  CMB data are in about $2.8 \sigma$ tension with the DESI DR2 data.  A more detailed discussion of this issue can be found in Section VII C ``Evaluating the consistency of CMB and BAO data in $\Lambda$CDM'' of the SPT data release  \cite{SPT-3G:2025bzu}. The authors advise caution in interpreting CMB+DESI data results in $\Lambda$CDM. For a general discussion of combining mutually inconsistent data sets, see e.g. \cite{LuisBernal:2018drn}.

A further investigation of this issue in \cite{Ferreira:2025lrd}  demonstrates that the fit of $\Lambda$CDM to CMB data exhibits a high degree of correlation
between $n_{s}$ and the BAO parameters $r_{d}h$ and $\Omega_{m}$. This correlation, and the tension in $r_{d}h$ and $\Omega_{m}$ between CMB data and DESI, can lead to significant shifts in $n_{s}$ when the two are combined. The authors concluded: ``Given the crucial role of $n_{s}$ in discriminating between inflationary models, we urge caution in interpreting CMB+BAO constraints on $n_{s}$ until the BAO-CMB tension is resolved.''

Instead of making a judgment call in this respect, we will consider in this paper the simplest inflationary models consistent with the Planck/BICEP/Keck results shown in Fig. \ref{BICEPKeck}, and then we will discuss the models matching the results of Planck, ACT, and DESI (P-ACT-LB)  shown in Fig. \ref{ACT}.

But before doing it, we will describe a basic chaotic inflation model with a polynomial potential with 3 parameters, which can describe any value of the amplitude of the spectrum of scalar perturbations $A_{s}$, of the spectral index $n_{s}$ and of the tensor-to-scalar ratio $r$.  This may help us view the rest of the paper from a different perspective.

\section{Polynomial chaotic inflation}\label{sec:polyn}

According to \cite{BICEP:2021xfz}, the simplest models of chaotic inflation $V(\phi) \sim \phi^{n}$ are ruled out by observations. However, it is not widely recognized that one can easily describe all presently available CMB-related data using models with simple polynomial potentials and only three parameters, as seen in Fig. 2 of \cite{Destri:2007pv}, Fig. 2 of \cite{Nakayama:2013jka}, and Fig. 3 of \cite{Kallosh:2014xwa}.

\begin{figure}[t!]
\begin{center}
\hskip -0.76 cm \includegraphics[scale=0.37]{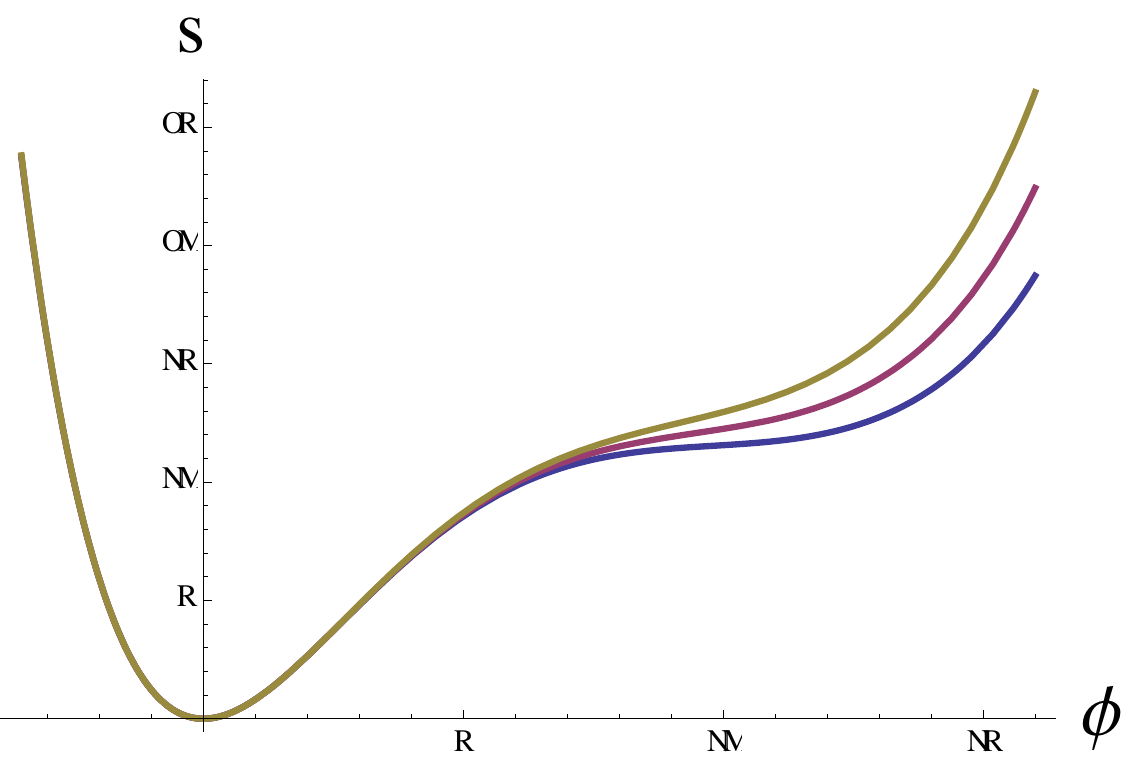}
\end{center}
\caption{\footnotesize The potential $V(\phi) = {m^{2}\phi^{2}\over 2}\,\bigl(1-a\phi +a^{2}b\,\phi^{2})\bigr)^{2}$ for $a = 0.1$ and $b = 0.36$ (upper curve), 0.35 (middle) and 0.34 (lower curve). The potential is shown in units of $m$, with $\phi$ in units $M_{p}= 1$. }
\label{chi}
\end{figure}

As a particular example, we will describe a model with the potential 
\be\label{3par}
V = {m^{2}\phi^{2}\over 2 } (1-a\phi + b (a\phi)^{2})^{2} \ .
\ee
This potential naturally appears in the simplest versions of chaotic inflation in supergravity  \cite{Kallosh:2014xwa}. For $ a= b = 0$, this potential coincides with the basic chaotic inflation potential ${m^{2}\phi^{2}\over 2 }$ \rf{chaot}.  The parameter $a$ stretches the potential horizontally and vertically while preserving its general shape. The parameter $b$ controls its asymptotic behavior at large $\phi$, see Fig. \ref{chi}. The values of $n_{s}$ and $r$ are determined by the parameters $a$ and $b$, and once they are fixed, one can adjust the amplitude of perturbations $A_{s}$ by changing $m$.

\begin{figure}[t!]
\begin{center}
 \includegraphics[scale=0.2]{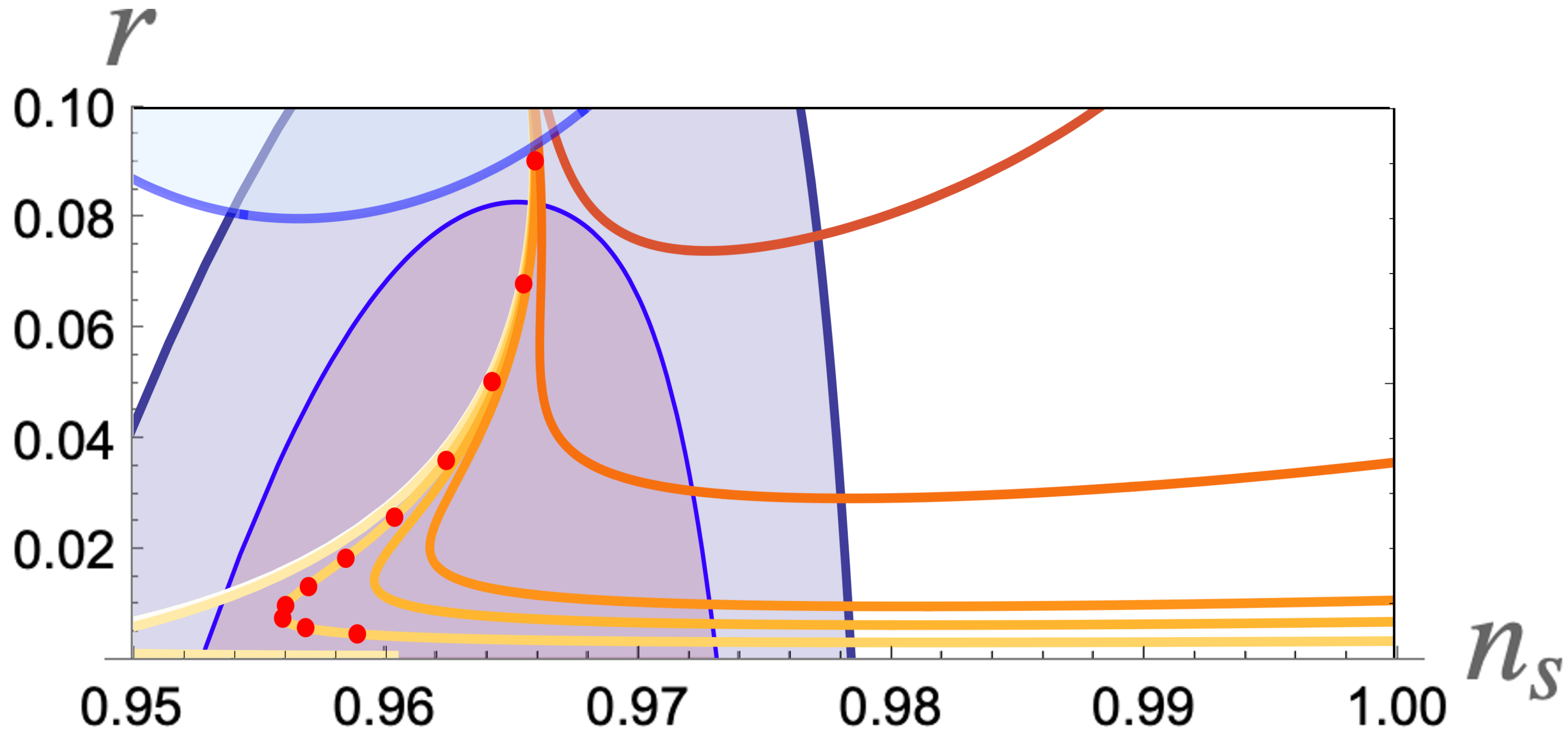}
\end{center}
\caption{\footnotesize Predictions for $n_{s}$ and $r$ at $N_{e} = 55$ in the model with $V(\phi) = {m^{2}\phi^{2}\over 2}\,\bigl(1-a\phi +a^{2}b\,\phi^{2})\bigr)^{2}$ for $b = 0.34$. The red dots, from the top down, correspond to $ a = 0$, 0.03, 0.05, 0.1, 0.13. For $a = 0$, one recovers the predictions for the simplest chaotic inflation model with a quadratic potential;  for $a = 0.13 - 0.14$, the predictions almost exactly coincide with the predictions of the Starobinsky model and the Higgs inflation model.}
\label{chi2}
\end{figure}

The values of $n_{s}$ and $r$  for $N_{e} = 55$ in this model are shown in Fig. \ref{chi2}, which is the lower part of Fig. 3 in \cite{Kallosh:2014xwa}. Let us consider, for example, the yellow line with many small red dots. This line corresponds to $b = 0.34$.  The red dots, from top to bottom, correspond to $a$ changing from $a = 0$ to $a = 0.13$. 

For completeness, we present the results for $N_{e} = 60$ here. For $b = 0.34$, $a = 0.13$, the predictions nearly coincide with the predictions of the Starobinsky model and the Higgs inflation model, as well as with the predictions of $\alpha$-attractors with $\alpha = 1$: \ $n_{s}= 0.967$, $r \approx 3\times 10^{{-3}}$. For $a = 0.14$, the value of $n_{s}$ becomes $0.973$, matching the ACT results shown in Fig. \ref{ACT}. A further increase of $a$ gradually increases the value of $n_{s}$, which reaches  $n_{s}= 1.006$ for  $a = 0.17$.

To adjust the value of $r$, it is sufficient to slightly change $b$, and then again slide along the corresponding nearly horizontal yellow line towards any desired value of $n_{s}$.   For example, for  $b = 0.35$ one has $r \approx 10^{{-2}}$. For $a = 0.09$ one has $n_{s}= 0.967$,  which grows to  $n_{s}= 1.003$ for $a = 0.13$. Meanwhile for $b = 0.335$ one has $r = 4\times 10^{-4}$. For $a = 0.2$ one has $n_{s}= 0.966$,  which grows up to $n_{s}= 1.01$ for $a = 0.25$.

Thus, we have a polynomial chaotic inflation model \rf{3par} with 3 parameters, $m$, $a$ and $b$, which can match any values of the 3 main inflationary parameters: the amplitude of scalar perturbations $A_{s}$,  the spectral index $n_{s}$, and the tensor-to-scalar ratio $r$. The problem of initial conditions for inflation in this model is trivially solved in the same way as in the simplest models of chaotic inflation. This model does not require any profound modifications of the fundamental theory. There is no need to consider modifications of the general theory of relativity, as in $R+R^2$  Starobinsky model, which we will discuss shortly. There is no need to invoke non-minimal coupling of the inflaton field to gravity  $\xi\phi^{2 }R$ with $\xi \gg 1$, as in the Higgs inflation, or modify its kinetic terms, as in the theory of $\alpha$-attractors. This version of inflationary theory just works.  It provides a basic framework that can describe {\it any}\, inflationary parameters  $ A_s$, $ n_s$, and $r$. In particular, a proper choice of the parameters $a$ and $b$ allows one to match any values of $n_{s}$ and $r$ favored by either Planck/BICEP/Keck or ACT.

Then why do we even need to continue discussing other models? The main reason is that some of them can match the  Planck/BICEP/Keck data using just a single parameter required to fix the amplitude of the perturbations $A_{s}$. Moreover, some of these models (attractors) have universal predictions that are very stable with respect to the choice of the inflationary potential. Finally, some of the models may have better motivation from the perspective of quantum gravity, supergravity, and/or string theory.  Such models, in addition to explaining  the data, may provide us with important insights into
fundamental physics.

\section{Starobinsky model}\label{sec:star}
In most of the recent papers, the authors use the name ``Starobinsky model'' to describe, interchangeably,  one of the two different models,
\be\label{starR}
{{\cal L} \over \sqrt {-g} } = 
 {R\over 2} + {R^{2}\over 12 M^{2}}  \ ,
\ee
or 
\be\label{starf}
{{\cal L} \over \sqrt {- \tilde g} }=  {  \tilde R\over 2} + {1\over 2}\partial_{\mu}\vp \partial^{\mu} \vp +{3M^{2}\over 4} \left(1-e^{-\sqrt{2\over 3} \vp}\right)^{2} \ .
\ee
However, as we will discuss now, the relation between these two models is somewhat subtle.  Moreover, the original version of the Starobinsky model \cite{Starobinsky:1980te} was slightly different.

 Instead of modifying the action, Starobinsky in its original model \cite{Starobinsky:1980te}  considered the  Einstein equations with quantum corrections to the energy-momentum tensor,
\begin{eqnarray}\label{Star1980}
\langle T_{ik} \rangle &=& {k_{2} \over 2880\pi^{2}} \Bigl(R_{i}^{l} R_{kl}- {2 \over  3} R R_{ik} - {1 \over 2} g_{ik} R_{lm}R^{lm} + {1 \over 4} g_{ik} R^{2}\Bigr) \\ \nonumber
& + & {k_{3} \over 2880\pi^{2}}\, {1 \over 3}\, \Bigl(R_{;i}^{;k} R_{kl} -    g_{ik} R^{;l}_{;l}  -  R R_{ik}  + {1 \over 4} g_{ik} R^{2}\Bigr) \ .
\end{eqnarray}
According to the original version of the Starobinsky model, the universe at $t = 0$ was in an unstable de Sitter state with the Hubble constant $H$ many orders of magnitude smaller than the Planck mass. 
In this sense, the original model was similar to the new inflation, with inflation beginning at the top of the potential.

Since de Sitter space is non-singular, one should be able to continue this solution to $t\to -\infty$. However, an unstable de Sitter state could not survive from $t = -\infty$ to $t = 0$. Another problem was finding a realistic model with proper parameter values $k_{2}$ and $k_{3}$, which proved very difficult.

The streamlined formulation of the Starobinsky model  \rf{starR} was proposed in 1985 \cite{Kofman:1985aw}, where Starobinsky explained why it was difficult to develop a realistic version of the original model  \cite{Starobinsky:1980te}. A transition from \rf{starR} to \rf{starf} took a few more years and the efforts of several authors \cite{Whitt:1984pd, Barrow:1988xh, Maeda:1987xf, Coule:1987wt}.

The Starobinsky model predicts \cite{Mukhanov:1981xt}
\be
\label{starpred}
 A_{s} = {V_{0} N_{e}^{2}\over 18 \pi^{2 }} \ , \qquad n_{s} = 1-{2\over N_{e}} \ , \qquad r = {12\over N^{2}_{e}} \ .
\ee 
where $V_{0} = {3M^{2}\over 4}$ is the height of the inflationary plateau in \rf{starf}.

\section {Higgs inflation and $\xi$-attractors}\label{sec:xi}
The basic idea of Higgs inflation was introduced long ago \cite{Salopek:1988qh}, but it was beautifully formulated and attracted lots of attention after the paper by Bezrukov and Shaposhnikov \cite{Bezrukov:2007ep}:
\be
{1\over \sqrt{-g}} \mathcal{L}=  \tfrac12 (1+\xi \phi^{2}) R  - \tfrac12 (\partial \phi)^2 -  \lambda^{2} \phi^{4}  \ . \label{Higgs} 
 \end{equation} 
In this model, the Higgs field is responsible for both inflation and the symmetry breaking in the Standard Model. This is a beautiful and very economical possibility. This model, in the large $\xi$ limit,  has the same predictions  $n_{s}= 1-{2\over N_{e}}$, $r = {12\over N^{2}_{e}}$ as the Starobinsky model, but because of the very efficient reheating in the Higgs inflation, the predicted numerical value of $n_{s}$ tends to be higher than the value of $n_{s}$ in the Starobinsky model.  For further developments of the Higgs inflation and its comparison with the Starobinsky model, see e.g. \cite{Barvinsky:2008ia,Bezrukov:2009db,Barvinsky:2009fy,Bezrukov:2010jz,Ferrara:2010yw,Ferrara:2010in,Bezrukov:2011gp,Karananas:2022byw}.
 
In general, both models match the Planck data very well, and the fact that these very different models have nearly identical predictions was enigmatic.  This prompted attempts to identify the reasons for this coincidence and to seek other models with similar properties. 

It was soon realized that there is a large class of models, $\xi$-attractors, which, in the large $\xi$ limit,  give the same predictions for $n_{s}$ and $r$  \cite{Kallosh:2013tua,Kallosh:2014rha}:
\be
{1\over \sqrt{-g}} \mathcal{L}=  \tfrac12 (1+\xi f(\phi)) R  - \tfrac12 (\partial \phi)^2 -  \lambda^{2} f^{2}(\phi)  \ . \label{JordanGen} 
 \end{equation} 
 Due to the non-minimal coupling of the scalar field to gravity, we will refer to this form of the theory as the Jordan frame. In order to transform to the canonical Einstein frame, we redefine the metric $  g_{\mu \nu} \rightarrow \Omega(\phi)^{-1} g_{\mu \nu}$.
This brings the Lagrangian to the Einstein-frame form:
 \begin{align}
 {\mathcal{L}_{\rm E}\over \sqrt{-g}} =  \tfrac12   R - \tfrac12 \Big (\Omega(\phi)^{-1} + \tfrac32 (\log \Omega(\phi))'^2\Big ) (\partial \phi)^2    
 - V_E(\phi)  \, , \quad {\rm where~~} V_E(\phi) =  \frac{V_J(\phi)}{\Omega(\phi)^2}  \,. \label{Einstein}
 \end{align}
The kinetic terms in \eqref{Einstein} give rise to the following definition of the canonical scalar field $\varphi$:
 \begin{align}
 \frac{\partial \varphi}{\partial \phi} = \sqrt{\Omega(\phi)^{-1} + \tfrac32 (\log \Omega(\phi))'^2} \,,
 \end{align}
 Many models discussed in this paper have non-minimal canonical terms. Sometimes the transition to the canonical variables $\vp$ is simple, sometimes it is not. A general method to investigate such models and to find  $n_{s}$ and $r$ beyond the slow-roll approximation can be found in Appendix \ref{App:A} where we describe this method for convenience.

\begin{figure}[H]
\centering
\includegraphics[scale=0.42]{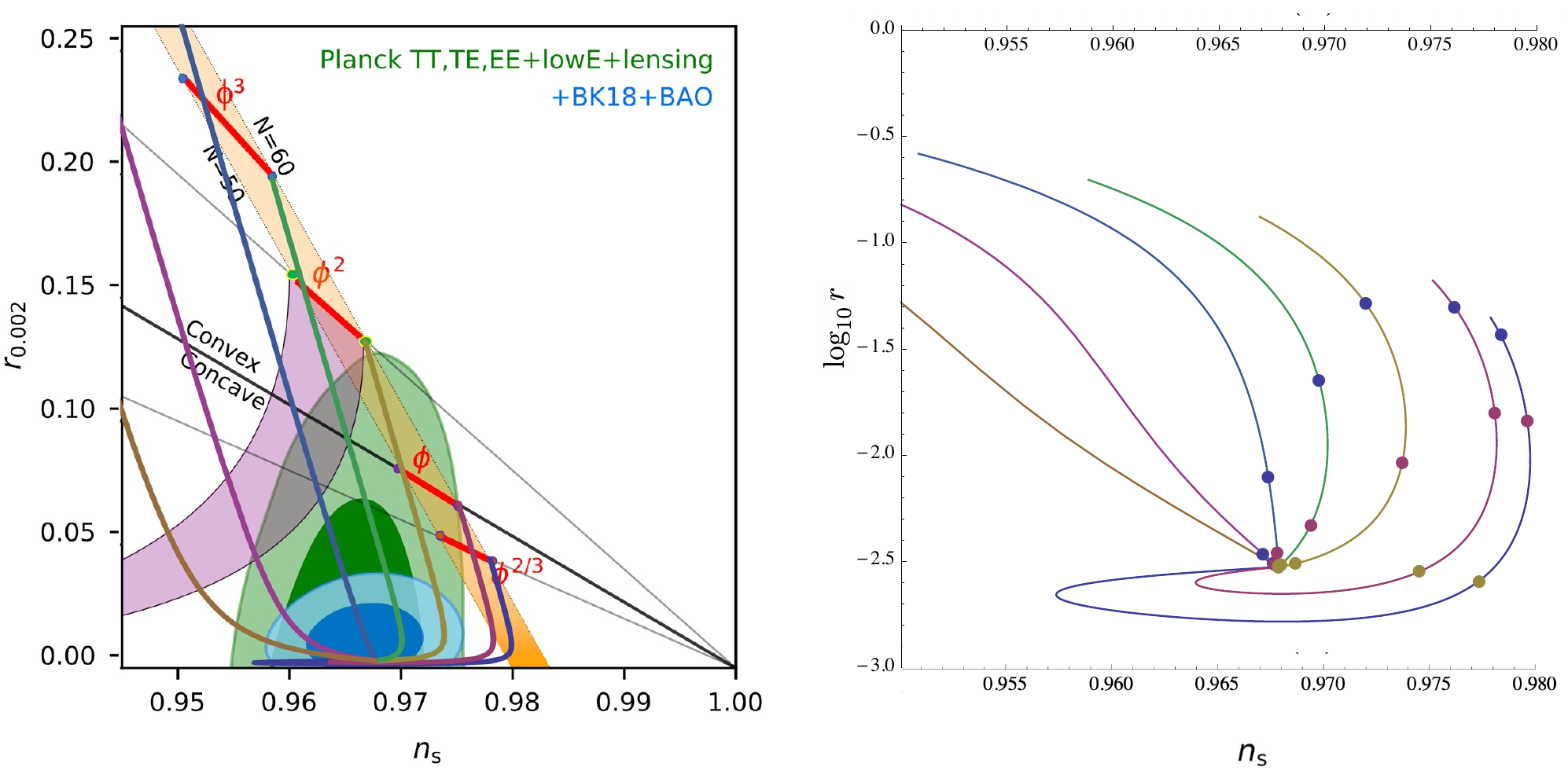}\vskip -5pt
\caption{\footnotesize  Attractor trajectories 
for  monomial models $V(\phi) \sim  f^{2}(\phi) \sim \phi^{8}, \phi^{6}, \phi^4, \phi^2, \phi, \phi^{2/3}$, for $N_{e}= 60$. The results of  \cite{Kallosh:2013tua} are superimposed with the BICEP/Keck results shown in Fig. \ref{BICEPKeck}. The dots in the right panel correspond to $\xi=1/e$, 1, and $e$, from the top down. These small details will be important for us when discussing the ACT data. }
\label{xi}
\end{figure}

The values of $n_{s}$ and $r$ as a function of $\xi$ are shown in Fig. \ref{xi} for $V(\phi) \sim \phi^{8}, \phi^{6}, \phi^4, \phi^2, \phi, \phi^{2/3}$. The dots in the right panel correspond to $\xi=1/e$, 1, and $e$, from the top down. As one can see, the convergence of the predictions for the potentials $V(\phi) \sim \phi^{8}, \phi^{6}, \phi^4$ to the attractor point  $n_{s}= 1-{2\over N_{e}}$, $r = {12\over N^{2}_{e}}$ is very fast, the attractor point is reached already at $\xi = O(1)$. Meanwhile for the models $\phi^2, \phi, \phi^{2/3}$ this limit is reached only for relatively large $\xi$. Thus, for these models, it makes sense to consider a more natural possibility $\xi = O(1)$. This consideration will be important for us later, see section \ref{nonmchaot}.  

 \section{Conformal attractors}\label{tmod}
 
 The second class of cosmological attractors leading to the same predictions as the Starobinsky model and Higgs inflation for a broad class of potentials has a rather unusual action \cite{Kallosh:2013hoa}
\begin{equation}
{\mathcal{L}\over \sqrt{-{g}}} =  {1\over 2}\partial_{\mu}\chi \partial^{\mu}\chi  +{ \chi^2\over 12}  R({g})- {1\over 2}\partial_{\mu} \phi\partial^{\mu} \phi   -{\phi^2\over 12}  R({g}) -{1\over 36} F\left({\phi/\chi}\right)(\phi^{2}-\chi^{2})^{2} \,.
\label{conform}
\end{equation}
where $F$ is an arbitrary function in terms of the homogeneous variable $z = \phi/\chi$. 

This theory is locally conformal invariant under the  transformations  
\be \tilde g_{\mu\nu} = e^{-2\sigma(x)} g_{\mu\nu}\,
,\qquad \tilde \chi =  e^{\sigma(x)} \chi\, ,\qquad \tilde \phi =  e^{\sigma(x)}
\phi\ . \label{conf}\ee 

The field $\chi(x)$ is referred to as a conformal compensator.  Its kinetic term has a negative sign, but this is not a problem because it
can be removed from the theory by fixing the gauge symmetry
(\ref{conf}). In particular, one can use the gauge $\chi^2-\phi^2=6$ and resolve this constraint in terms of the fields 
$\chi=\sqrt 6 \cosh  {\varphi\over \sqrt 6}$,  $ \phi= \sqrt 6 \sinh {\varphi\over \sqrt 6} $ 
and the   field $\varphi$:~ ${\phi\over\chi} = \tanh{\varphi\over \sqrt 6}$. 
Our action \rf{conform} becomes
\begin{equation}\label{chaotmodel}
{\mathcal{L}\over \sqrt{-{g}}} =  \frac{1}{2}R - \frac{1}{2}\partial_\mu \varphi \partial^{\mu} \varphi -   F(\tanh{\varphi\over \sqrt 6})  \ .
\end{equation}
Note that $\tanh\varphi\rightarrow \pm 1$ in the large $\varphi$ limit, and therefore $F(\tanh{\varphi\over \sqrt 6})  \rightarrow \rm const$.  This class of models has the same general predictions $n_{s}= 1-{2\over N_{e}}$, $r = {12\over N^{2}_{e}}$ as the Starobinsky model and the Higgs inflation model, nearly independently of the choice of the function $F$   \cite{Kallosh:2013hoa}.
This class of models was the last step towards the development of $\alpha$-attractors.

\section{\boldmath{$\alpha$-attractors}}\label{sattr}

The theory of $\alpha$-attractors \cite{Kallosh:2013hoa,Ferrara:2013rsa,Kallosh:2013yoa,Galante:2014ifa,Kallosh:2015zsa,Kallosh:2019eeu,Kallosh:2019hzo} has its roots in the hyperbolic geometry of the moduli space \cite{Kallosh:2015zsa}, and in supergravity \cite{Kallosh:2025jsb}. Here we will describe the phenomenological implications of two basic classes of single-field $\alpha$-attractors, T-models and E-models.

\subsection{T-models}\label{Sec:Tmodels}
The simplest T-model is given by 
 \be
{ {\cal L} \over \sqrt{-g}} =  {R\over 2}  -  {(\partial_{\mu} \phi)^2\over 2\bigl(1-{\phi^{2}\over 6\alpha}\bigr)^{2}} - V(\phi)   \,  .
\label{cosmoA}\ee
Here $\phi(x)$ is the inflaton.  In the limit $\alpha \to \infty$, the kinetic term becomes the canonical term $-  {(\partial_{\mu} \phi)^2\over 2}$, as in the original chaotic inflation model \rf{chaot}.  One can recover the canonical normalization for any $\alpha$ by solving the equation ${\partial \phi\over 1-{\phi^{2}\over 6\alpha}} = \partial\vp$. The solution is $
\phi = \sqrt {6 \alpha}\, \tanh{\varphi\over\sqrt {6 \alpha}}$.
The full theory, in terms of the canonical variables  $\vp$,  becomes a theory with a plateau potential
 \be
{ {\cal L} \over \sqrt{-g}} =  {R\over 2}  -  {(\partial_{\mu}\varphi)^{2} \over 2}  - V\big(\sqrt {6 \alpha}\, \tanh{\varphi\over\sqrt {6 \alpha}}\big)   \,  .
\label{cosmoqq}\ee
In particular, for $V(\phi) = {V_{0}\over (6 \alpha)^{n}} \phi^{2n}$ one finds
\be\label{Tmodel}
V(\phi) = V_{0} \tanh^{2n}{\varphi\over\sqrt {6 \alpha}} \ .
\ee
At large $\vp$, the potential can be represented as 
\be\label{plateau1}
V(\vp) = V_{0} - 2  \sqrt{6\alpha}\,V'_{0} \ e^{-\sqrt{2\over 3\alpha} \varphi } \ .
\ee
Here $V_0 = V(\phi)|_{\phi =  \sqrt {6 \alpha}}$ is the height of the plateau potential, and $V'_{0} = \partial_{\phi}V |_{\phi = \sqrt {6 \alpha}}$. The coefficient $2  \sqrt{6\alpha}\,V'_{0}$ in front of the exponent can be absorbed into a shift of the field $\varphi$. 

That is why all inflationary predictions in the regime with $e^{-\sqrt{2\over 3\alpha} \varphi } \ll 1$ are determined by two parameters, $V_{0}$ and $\alpha$,  and not by any other features of the potential $V(\phi)$:
\be
\label{pred}
 A_{s} = {V_{0}\, N_{e}^{2}\over 18 \pi^{2 }\alpha} \ , \qquad n_{s} = 1-{2\over N_{e}} \ , \qquad r = {12\alpha\over N^{2}_{e}} \ .
\ee 
Accuracy of these results increases with an increase of $N_{e}$ and a decrease of $\alpha$. These results are compatible with all presently available Planck/BICEP/Keck data. 
\begin{figure}[H]
\centering
\includegraphics[scale=0.29]{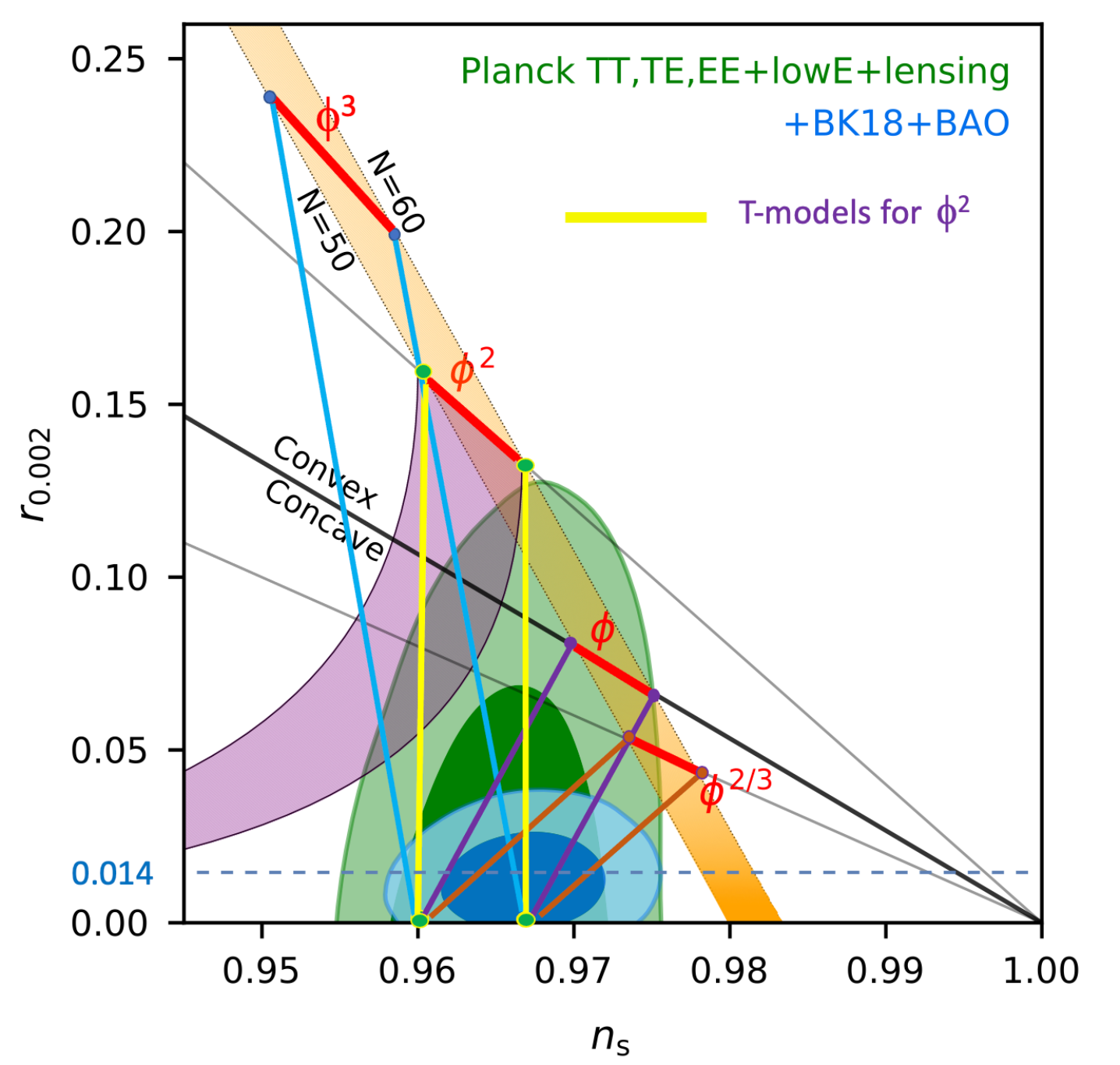}
\vskip -10pt
\caption{\footnotesize The figure illustrating the main results of the BICEP/Keck \cite{BICEP:2021xfz} superimposed with the predictions of  $\alpha$-attractor T-models with the potential $\tanh^{2n} {\varphi\over \sqrt{6\alpha}}$ \cite{Kallosh:2013yoa,Kallosh:2015zsa}. In the large $\alpha$ limit, $r$ predictions of each of these models coincide with the predictions of the models  $\phi^{2n}$, but for $\alpha \ll 1$ these predictions converge in the dark blue area favored by the BICEP/Keck \cite{Kallosh:2013yoa}. 
}
\label{Fan}
\end{figure}
To illustrate this result, we show in Fig. \ref{Fan} predictions of the models with several monomial potentials after the modification of the kinetic term shown in \rf{cosmoA}. At large $\alpha$, predictions of all of these models coincide with the predictions shown in Fig. \ref{BICEPKeck0},  but at  $\alpha \lesssim 1$ their predictions converge to the dark blue area favored by the latest BICEP/Keck data release.  Fig. \ref{Fan} illustrates the main advantage of the cosmological attractors.  

The amplitude of inflationary perturbations in these models matches the Planck normalization $A_{s} \approx 2.01 \times 10^{{-9}}$ for  $ {V_{0}\over  \alpha} \sim 10^{{-10}}$, $N_{e} = 60$, or for $ {V_{0}\over  \alpha} \sim 1.5 \times 10^{{-10}}$, $N_{e} = 50$.  For the simplest  model $V = {m^{2}\over 2} \phi^{2}$  one finds
\be\label{T}
V =  3m^{2 }\alpha \tanh^{2}{\varphi\over\sqrt {6 \alpha}} \ .
\ee
This simplest model is illustrated by the prominent vertical yellow band in Fig. 8 of the paper on inflation in the Planck 2018 data release \cite{Planck:2018jri}.  In this model,  the condition $ {V_{0}\over  \alpha} \sim 10^{{-10}}$ reads $ m  \sim   0.6 \times10^{{-5}}$. The small magnitude of this parameter accounts for the small amplitude of perturbations $A_{s} \approx 2.01 \times 10^{{-9}}$. No other parameters are required to match the presently existing Planck/BICEP/Keck data in this model. If the inflationary gravitational waves are discovered, their amplitude can be accounted for by the choice of the parameter $\alpha$ in \rf{pred}.
\subsection{E-models}
\begin{figure}[H]
\vskip -5pt
\centering
\includegraphics[scale=0.36]{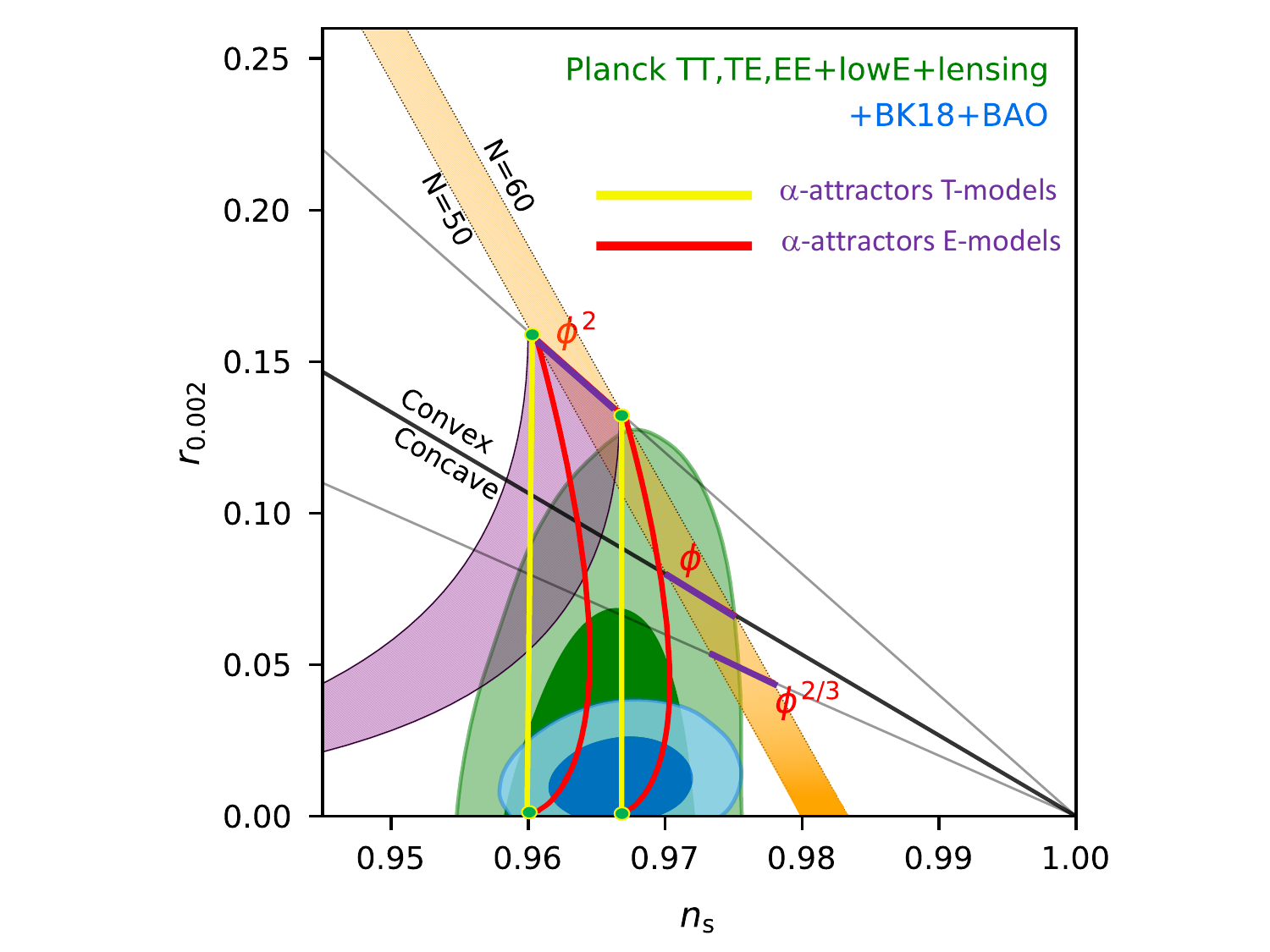}
\vskip -15pt
\caption{\footnotesize  BICEP/Keck results for $n_{s}$ and $r$ \cite{BICEP:2021xfz} superimposed with the predictions of the simplest $\alpha$-attractor T-model with the potential $\tanh^{2} {\varphi\over \sqrt{6\alpha}}$ (yellow lines for $N_{e} = 50, 60$) and E-models  with the potential $\big (1-e^{-\sqrt{{2\over 3\alpha}}\varphi}\big )^{2}$ (red lines for $N_{e} = 50, 60$).}
\label{BICEPKeck0}
\end{figure}
The second family of $\alpha$-attractors, called E-models, is given by
 \be\label{actionE}
{ {\cal L} \over \sqrt{-g}} =  {R\over 2} - {3\alpha\over 4} \, {(\partial \rho)^2\over \rho^{2}}- V(\rho) \ .
\ee
As before, one can go to canonical variables,  which yield
\be\label{apole}
{ {\cal L} \over \sqrt{-g}} =  {R\over 2} - {1\over 2}  (\partial \varphi)^2- V(e^{-\sqrt {2\over 3\alpha}\varphi}). 
\ee
We consider $V(\rho)$  not singular at $\rho = 0$, e.g. $V(\rho) =V_{0}(1-\rho)^{2n}$. In  canonical variables, it gives
\be\label{Emodel}
 V = V_0 \Bigl(1 - e^{-\sqrt {2 \over 3\alpha}\varphi} \Bigr)^{2n} . 
\ee
For $\alpha = 1$  and $n = 1$, this potential coincides with the potential of the Starobinsky model  \cite{Starobinsky:1980te}. In the small $\alpha$ limit, the predictions of the E-models coincide with the predictions of the T-models \rf{pred}, see Fig. \ref{BICEPKeck0}.
\begin{figure}[H]
\centering
\includegraphics[width=140mm]{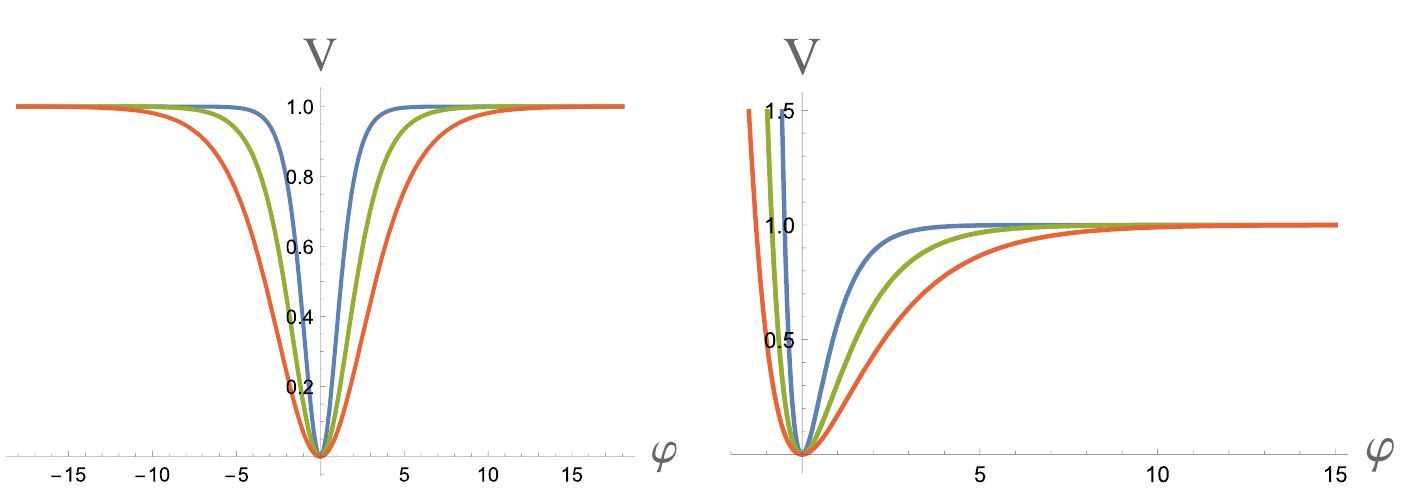}
\caption{\small On the left panel, we show the potential of the  T-model   \rf{Tmodel} for $n=1$, $V_0=1$, $3\a=1,3,7$. On the right, there is an E-model potential  \rf{Emodel} for the same parameters. The T-model potential with $\a=1, n=1$ is very similar to the potential of the Higgs inflation. The E-model potential with  $\a=1, n=1$ coincides with the potential in the Starobinsky model.}
\label{fig:TT}
\end{figure}
It is interesting that the red curves describing E-models at $0.02\lesssim r  \lesssim 0.03$ in Fig. \ref{fig:TT} bend to the right and reach $n_s\approx 0.97$ for $N_{e} = 60$, in agreement with  Planck and ACT  (P-ACT) result $n_{s} \approx  0.97$. As we will see later in Fig. \ref{TE}, the red line touches the 1$\sigma$ region of DESI (P-ACT-LB) data $n_{s} \approx 0.974  $  \cite{Louis:2025tst}, see section \ref{sec:TE}. The significance of this observation is in the fact that near-future B-mode experiments like BICEP-Keck and the Simons Observatory might probe this range of $r$. 

In Fig.  \ref{fig:TT} we show potentials of T-models and E-models for various values of their parameters.

\subsection{\boldmath Special cases: Poincare disks and $SL(2,\mathbb{Z})$ $\alpha$-attractors}\label{spec}
So far, we presented T- and E-models with a continuous range of values of $\alpha$. One can implement these models in the minimal ${\cal N}=1$ supergravity, where the parameter $3\alpha$  is given by
 ${2\over 3\alpha}= |{\cal R}_K|$. Here $|{\cal R}_K|$ is the curvature of  \K\, geometry \cite{Ferrara:2013rsa}.
In the context of the Poincar\'e hyperbolic disk geometry, representing an Escher disk,
$R^2_{\rm Esher} =3\alpha$  defines the size of the disk \cite{Kallosh:2015zsa}.
\begin{figure}[H]
\centering
\includegraphics[scale=0.3]{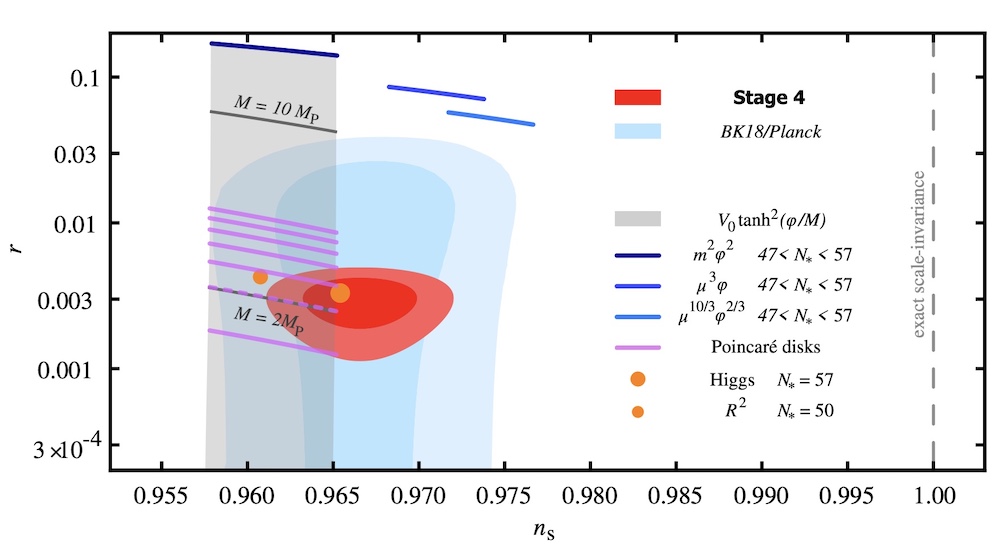}
\caption{\footnotesize The figure from  ``Snowmass2021 Cosmic Frontier: Cosmic Microwave Background Measurements White Paper'' \cite{Chang:2022tzj}. It shows the predictions of T-model $\alpha$-attractors with unconstrained values of $\alpha$ (gray area),  the predictions for $3\alpha= 7,6,5,4,3,2,1$ (purple lines), as well as Higgs inflation, $R^2$ inflation (red dots).  The predictions are for $47 < N_e< 57$.}
\label{Flauger0}
\end{figure} 
The most interesting B-mode targets in this class of cosmological attractor models are the ones with the discrete values of $3\alpha= 7,6,5,4,3,2,1$ 
\cite{Ferrara:2016fwe,Kallosh:2017ced,Gunaydin:2020ric,Kallosh:2021vcf}. These models of Poincar\'e disks are inspired by string theory, M-theory, and maximal supergravity. Predictions of these models are shown by purple lines in Fig. \ref{Flauger0}. 

In this context, we should also mention one of the more developed models of string theory inflation, called fibre inflation, see  \cite{Cicoli:2024bxw} and references therein. It was observed in \cite{Kallosh:2017wku}  that from the phenomenological point of view, fibre inflation
can be effectively described as an $\alpha$-attractor with  $\a=2$. This was also confirmed in  \cite{Cicoli:2024bxw}.
Therefore, fibre inflation predicts the values of  $n_s$ and $r$ shown in Fig. \ref{Flauger0} by the second from the top purple line corresponding to a Poincar\'e disk with $3\a=6$.

\begin{figure}[H]
\centering
\vskip -10pt
\includegraphics[scale=0.52]{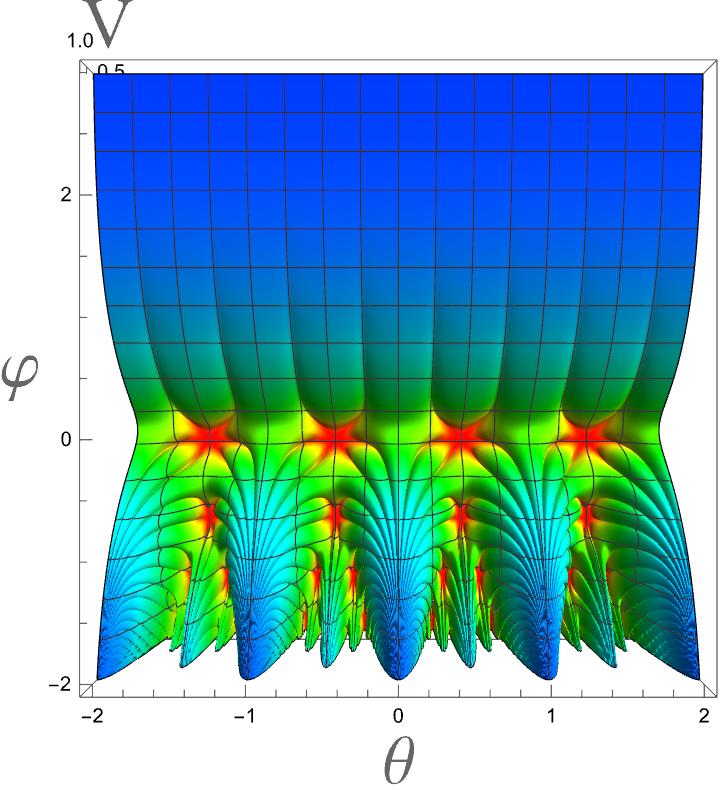}  
\caption{\footnotesize A typical $SL(2,\mathbb{Z})$ invariant potential in Cartesian coordinates \cite{Carrasco:2025rud}. The upper part of the figure shows an infinitely large inflationary plateau. The lower part of the figure shows an infinite number of sharp ridges. }
\label{cart}
\end{figure}

Recently, there were some interesting developments in the theory of $SL(2,\mathbb{Z})$ invariant $\alpha$-attractors \cite{Casas:2024jbw,Kallosh:2024ymt,Kallosh:2024pat,Kallosh:2024kgt,Carrasco:2025rud,GonzalezQuaglia:2025qem}. Potentials of these models have an infinitely large inflationary plateau, which has the same shape in the inflaton direction as the potentials of $\alpha$-attractors. But these potentials also have an infinite number of ridges, which, at first glance, appear too sharp to support inflation (see Fig. \ref{cart}). 

However,  this apparent sharpness is just an illusion created by hyperbolic geometry. One can show that each of these ridges is physically equivalent to an infinite inflationary plateau. Thus, these potentials describe an infinitely large number of inflationary plateaus. 

\begin{figure}[H]
\centering
\vskip -10pt
\includegraphics[scale=0.15]{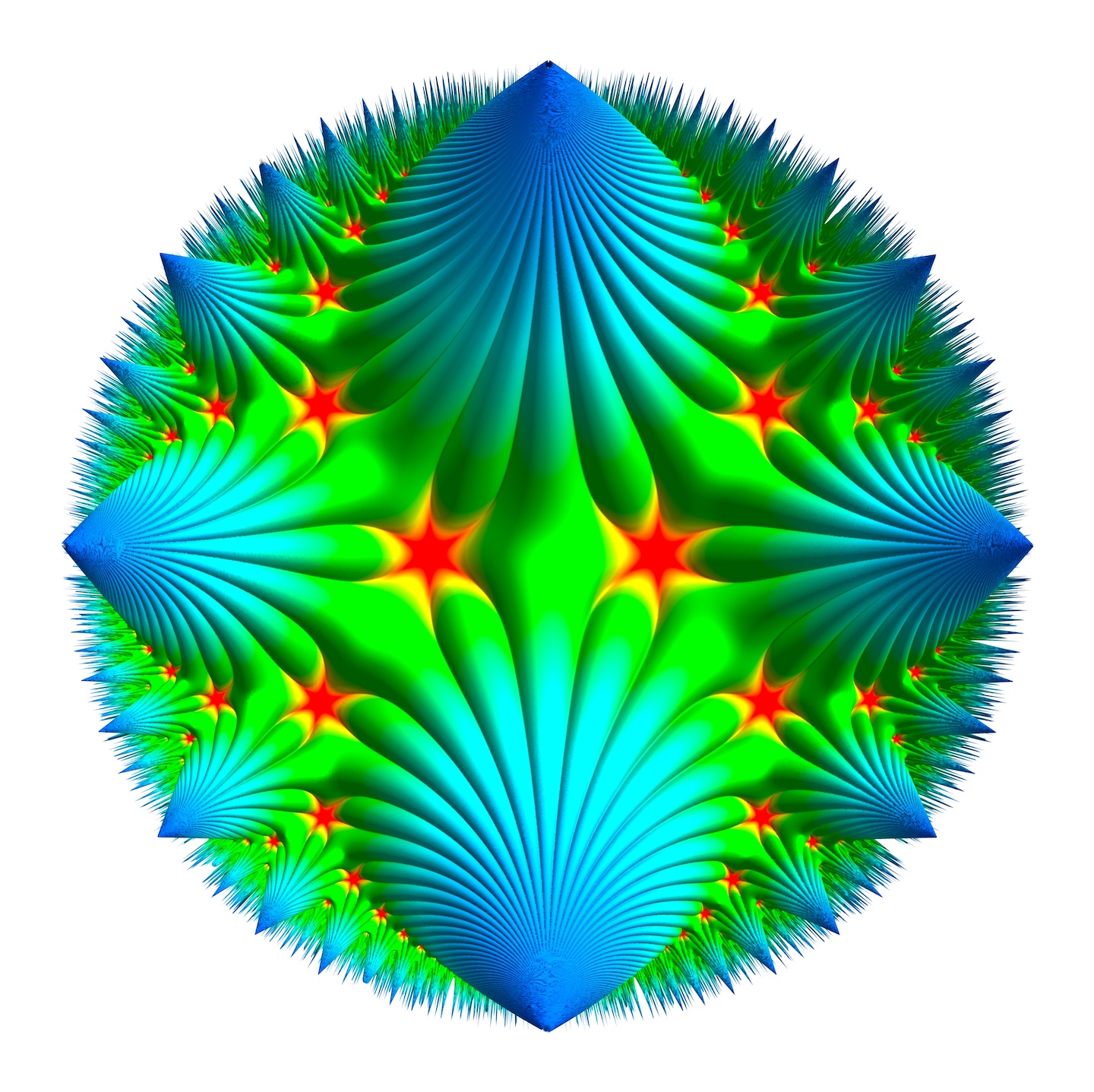} 
\caption{\footnotesize A typical $SL(2,\mathbb{Z})$ invariant potential in disk variables described on \cite{Carrasco:2025rud}.  }
\label{disk1000}
\end{figure}

The intricate structure of the inflationary potential in the models with axion stabilization is shown in Fig. \ref{disk1000} \cite{Carrasco:2025rud}. The sharp peaks in this figure correspond to large inflationary plateaus that are physically equivalent. All stabilized inflationary trajectories are shown by infinitely many axion valleys (dark lines) in the figure. Inflation occurs when the inflaton field rolls down along these valleys. 

Thus, the plot of the potentials with axion stabilization in disk variables immediately provides a detailed and complete map of the $SL(2,\mathbb{Z})$- invariant potentials, including a map of all inflationary trajectories available in these theories. After the axion stabilization \cite{Carrasco:2025rud}, these models have the same (discrete) set of predictions as the $\alpha$-attractor Poincar\'e disk models shown by the purple lines in Fig. \ref{Flauger0}.

\section{\boldmath ACT  and pre-attractor regime}\label{sec:TE}

As mentioned in Sect. \ref{constr}, the value of $n_s$  based on a combination of Planck, ACT, BICEP/Keck, and DESI is
\be
n_{s} =0.9743 \pm  0.0034
\label{ACT2}\ee
This shift to the right from the Planck 2018 constraint   $n_{s } = 0.9651 \pm  0.0044$ disfavors many models described in Sections \ref{sec:star}, \ref{sec:xi}, \ref{tmod}, and \ref{sattr}. This includes the Starobinsky model, the Higgs inflation model, many versions of $\xi$-attractors, and all versions of $\alpha$-attractor models predicting $n_{s} = 1-2/N_{e}$. If this new range of $n_{s}$ is confirmed by further investigations,  this  disfavors, to various degrees,   all  \ CMB-S4 and LiteBIRD 
targets shown in Fig. \ref{Flauger0}.
 \begin{figure}[H]
\begin{center}
\includegraphics[scale=0.3]{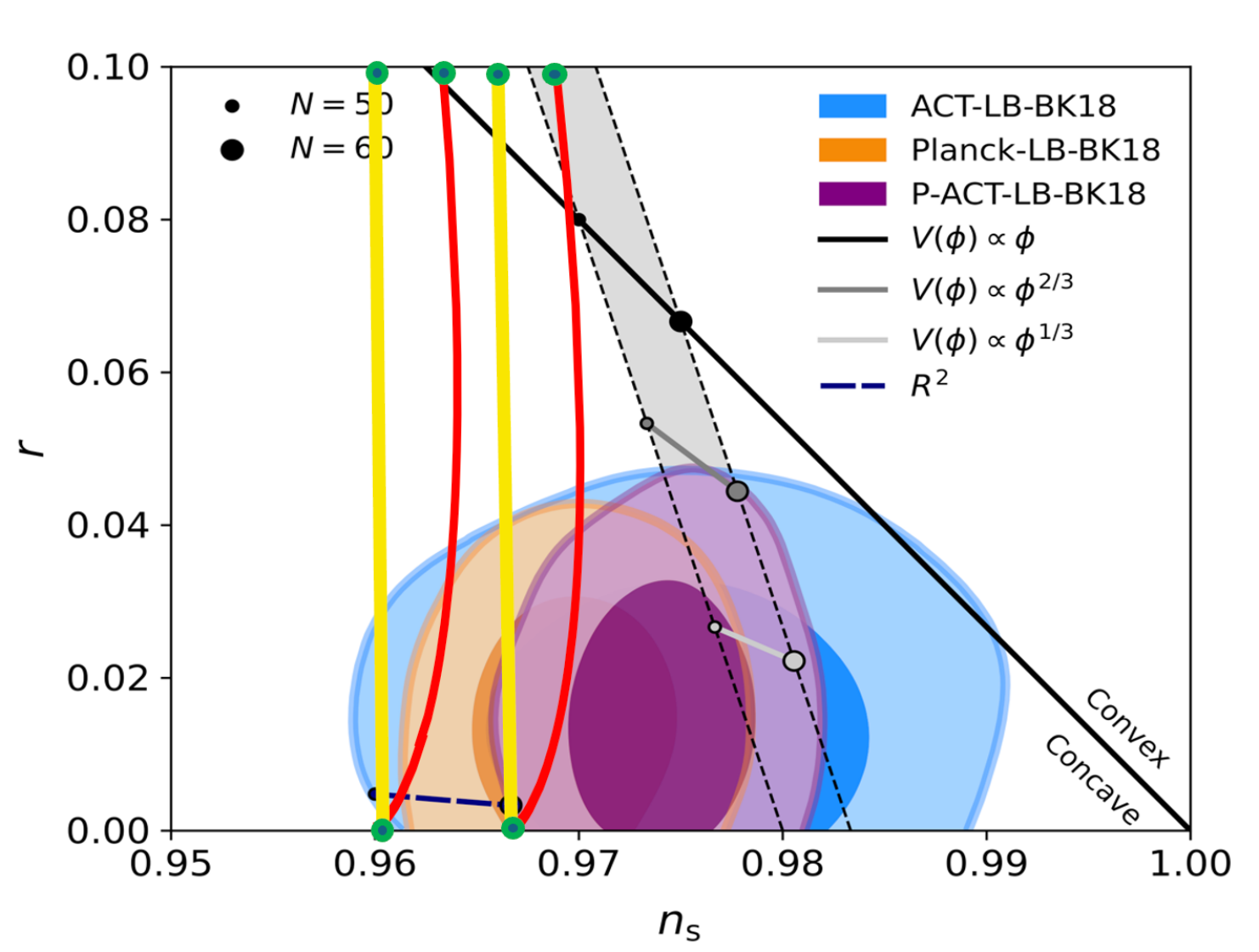}
\end{center}
\caption{\footnotesize ACT-Planck-LB-BICEP/Keck results for $n_{s}$ and $r$  superimposed with the predictions of the simplest $\alpha$-attractor T-model with the potential $\tanh^{2} {\varphi\over \sqrt{6\alpha}}$ (yellow lines for $N_{e} = 50, 60$) with the E-models $\big (1-e^{-\sqrt{{2\over 3\alpha}}\varphi}\big )^{2}$ (red lines for $N_{e} = 50, 60$)}
\label{TE}
\end{figure} 
At first glance, all of these models must be affected equally since all of them have the same attractor prediction $n_{s} = 1-2/N_{e}$. However, as we have already mentioned, typically reheating in the Starobinsky model is less efficient than in the Higgs inflation model, which is why $n_{s}$ in the Starobinsky model is expected to be smaller than in the Higgs inflation model, as shown in Fig. \ref{Flauger0}.

Also, a quick glance at Fig. \ref{TE}   shows that the predictions of the simplest T-models (yellow lines) and the simplest E-models (red lines) coincide in the attractor limit $r \ll 0.01$, but they are somewhat different in the pre-attractor regime at $0.015\lesssim r \lesssim 0.036$ corresponding to the dark purple area favored by ACT. In this range of $r$, the red curve with $N_{e} = 60$ almost touches the dark purple area favored by ACT, as shown in Fig. \ref{TE}. Thus, E-models with $r$ in this range are not incompatible with ACT. For a related discussion of this issue, see also \cite{Haque:2025uri}.
 
\begin{figure}[H]
\begin{center}
\includegraphics[scale=0.3]{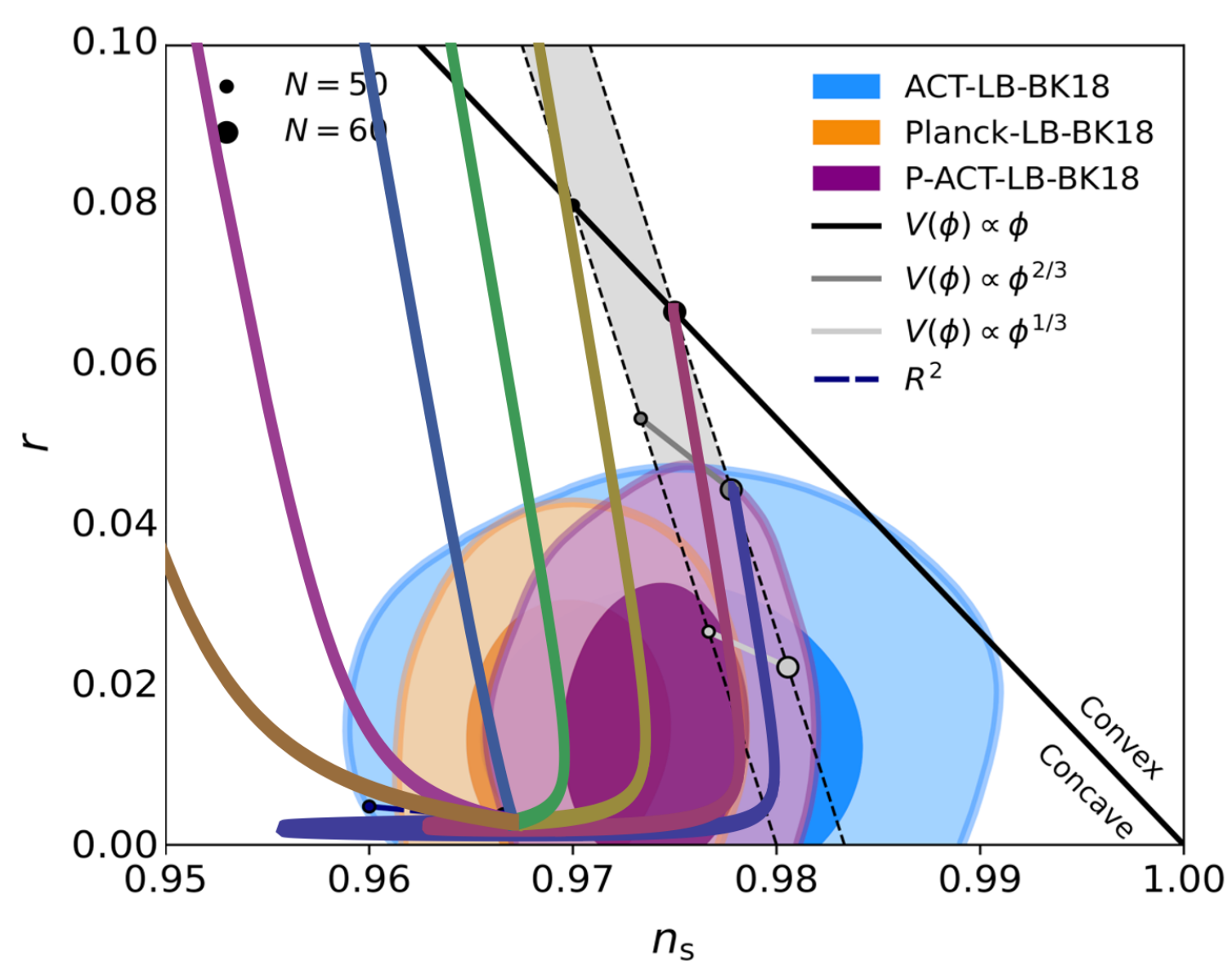}
\end{center}
\caption{\footnotesize Attractor trajectories 
for  $V(\phi) \sim  f^{2}(\phi) \sim \phi^{8}, \phi^{6}, \phi^4, \phi^3, \phi^2, \phi, \phi^{2/3}$, left to rught, for $N_{e}= 60$. The results of  \cite{Kallosh:2013tua} are superimposed with the ACT results shown in Fig. \ref{ACT}. The predictions of the model $V\sim \phi^2$ cross the central part of the dark purple area favored by ACT for $N_{e}= 60$, $0.3< \xi <4$. One can show that the curve corresponding to predictions for the linear potential $V\sim \phi$ crosses the central part of this area for $N_{e}= 50$. }
\label{xiact2}
\end{figure}
Another, even more striking example of the pre-attractor regime is given by $\chi$-attractors. As one can see in Fig. \ref{xi}, the attractor trajectories on their way towards the attractor point $n_{s} = 1-{2\over N_{e}},  r = {12\over N^{2}_{e}}$  span a wide range of values of $n_{s}$ while gradually descending to its attractor value, see Fig. \ref{xiact2}, where we superimpose these trajectories with the ACT results.   We will return to a more detailed discussion of this regime in subsection \ref{nonmchaot}. 
 In the next two sections, we will describe two other $\alpha$-attractor models, quintessential $\alpha$-attractors,  
where $n_{s}$ can approach the range \rf{ACT2}, 
and hybrid $\alpha$-attractors, which describe two-field models where $n_{s}$ can take {\it any}\, values from  $n_{s} = 1-2/N_{e}$ to $n_{s} =1$.


\section{\boldmath Quintessential $\a$-attractors, reheating  and ACT, DES, and DESI}\label{sec:reh}

It has been observed recently in \cite{Drees:2025ngb} that the predictions for the spectral index and tensor-to-scalar ratio in the Starobinsky model can lie within 1$\sigma$ of the recent ACT constraints if the reheating temperature satisfies 4\, MeV$ \leq T_{rh} \leq 10$  GeV and if equation of state during a long reheating process is in the range $0.8\leq w\leq 1$. Related results in application to the Starobinsky model and Higgs inflation can also be found in  \cite{Liu:2025qca,Haque:2025uri}. However, it was noted in \cite{Drees:2025ngb} that constructing a viable theory of this type is a challenge.

Fortunately, $\alpha$-attractors allow much greater flexibility in this respect. As an illustration, we will consider here two-shoulder quintessential $\alpha$-attractors. The potential depends on a geometric field $\phi$ related to a canonical field $\vp$ as $
\phi = \sqrt {6 \alpha}\, \tanh{\varphi\over\sqrt {6 \alpha}}$,  as explained in Sec. \ref{Sec:Tmodels}.
 The simplest example of a quintessential $\alpha$-attractor is a T-model with the linear dark energy potential  
\be\label{llll}
V(\phi) =  \gamma  \phi + \Lambda\, .
\ee
This was one of the first models of dark energy proposed back in 1986, and it simultaneously provided a simple anthropic solution to the cosmological constant problem  \cite{Linde:1986dq}. But the original version of this model required an exponentially small value of the parameter $\gamma$. This is no longer a problem in the context of $\alpha$-attractors  \cite{Akrami:2017cir}.

In terms of the canonically normalized field $\vp$, the linear potential \rf{llll} becomes
\be\label{linlin1}
V(\vp) = \gamma \sqrt{6\alpha}  (\tanh {\varphi\over\sqrt {6 \alpha}}+1)  + \Lambda\, .
\ee
\begin{figure}[h!]
\begin{center}
\includegraphics[scale=0.5]{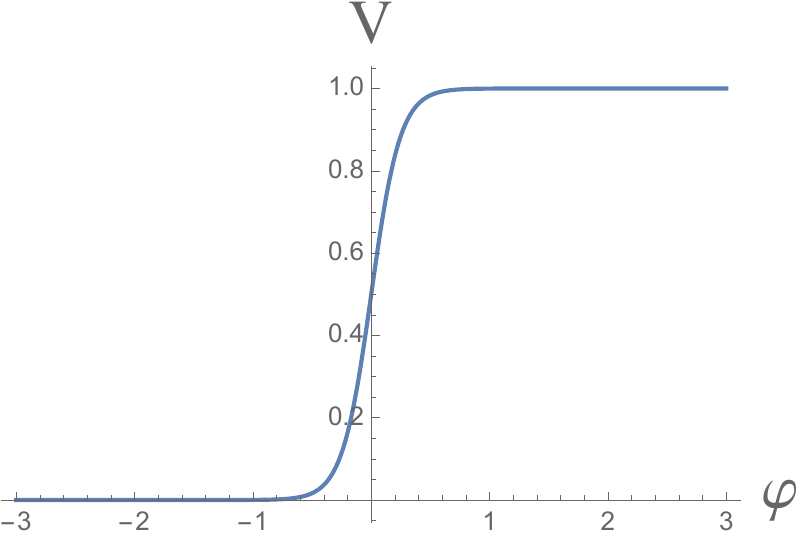}
\end{center}
\caption{\footnotesize Two-shoulder linear potential $V = {1\over 2\sqrt{6\alpha}}(\sqrt{6\alpha}+\phi) +\Lambda = {1\over 2} (1 + \tanh{\vp\over \sqrt{6\alpha}})+\Lambda$ for $\alpha = 10^{-2}$ and $\Lambda\sim 10^{{-120}}$.  The tiny cosmological constant $\Lambda$ is crucial for the validity of our scenario, but  $\Lambda$ is so small that it is invisible in this figure.}
\label{lin1}
\end{figure}
At  $\varphi \gg + \sqrt {6\alpha}$ and $\Lambda \ll \gamma \sqrt{6\alpha}$ this is the standard inflationary $\alpha$-attractor potential \be\label{pos2}
V(\vp)=   2\gamma \sqrt{6\alpha}(1 - e^{-\sqrt{2\over 3\alpha} \varphi })\, ,
\ee
whereas at   $\varphi \ll - \sqrt {6\alpha}$ it is a dark energy potential, asymptotically approaching a cosmological constant $\Lambda$, which can be positive, negative, or zero:
\be\label{linlin}
V(\vp)=   \Lambda + 2\gamma \sqrt{6\alpha}\  e^{\sqrt{2\over 3\alpha} \varphi } \, .
\ee
To become realistic, this model should have $\alpha \lesssim 0.02$ \cite{Akrami:2017cir}. This may not seem very natural, but the possibility of consistently describing inflation and dark energy using such a simple model is quite remarkable. Quintessential $\alpha$-attractors which may work for larger values of $\alpha$ have been introduced in \cite{Dimopoulos:2017zvq,Akrami:2017cir,Dimopoulos:2017tud}.

Many of these models, including the one shown in Fig. \ref{lin1}, have a potential that does not have a minimum. The standard reheating mechanism, in which an oscillating inflaton decays into other particles, does not work in these models. There are two basic mechanisms responsible for reheating in quintessential inflation: gravitational reheating \cite{Peebles:1998qn} and instant preheating \cite{Felder:1998vq,Felder:1999pv}. Gravitational reheating is very inefficient;  it may take a lot of time  \cite{Dimopoulos:2017zvq}, but one can increase its efficiency by taking into account instant preheating, thus effectively controlling the density of the universe at the end of reheating $\rho_{\rm reh}$, as well as the reheating temperature $T_{\rm reh}$ \cite{Akrami:2017cir}. 
Quintessential
$\alpha$-attractor models studied in \cite{Akrami:2017cir}, as well as in the earlier models in \cite{Linder:2015qxa,Dimopoulos:2017zvq,Dimopoulos:2017tud}, had the main purpose of describing evolving dark energy instead of the $\Lambda$CDM model, as in the general case of quintessential inflation  \cite{Peebles:1998qn}.
 It has been noticed in \cite{Dimopoulos:2017zvq,Akrami:2017cir} that the energy density of the inflaton during the lengthy stage of reheating in these models is dominated not by its potential but by its kinetic energy (the stage of kination). Therefore the parameter $w$ in the equation of stage $p = w \rho$  can be very close to $1$, 
  \be
 w\approx1 \ .
 \ee
The standard relation between the number of e-foldings $N_e$ and $w$ has a term of the form
 \be
 N_e= \dots + {1-3w\over 12(1+w)} \ln \Big({\rho_{\rm reh}\over \rho_{\rm end}}\Big) \ ,
 \ee
where $\rho_{\rm end}$ is the density of the universe at the end of inflation. This term is negative for $w = 0$ (dust), which is the regime encountered in the standard mechanism of reheating during oscillations. However, it becomes positive for $w \sim 1$ (kinetion).  It was shown in   \cite{Akrami:2017cir} that this can lead to a significant difference  $\Delta N_{e}$ between $N_e^{\rm osc}$  in the usual inflationary models with reheating during oscillations, and $N_e^{\rm quint}$ in the quintessential  $\a$-attractor models with a long stage of kination ending due to a combination of the gravitational reheating and instant preheating. In some cases studied in  \cite{Akrami:2017cir} $N_e^{\rm quint}$ can be greater than $N_e^{\rm osc}$ by about 10.

 The exact values of $N_e^{\rm quint}$ and, therefore, $n_s^{\rm quint}$, depend on the details of the model. But if due to this effect the number of e-foldings can increase from 60 to 70, this leads to an increase of $ n_s $ by about $0.006$. For example, if we take a typical $\alpha$-attractor model with  $N_{e} = 60 $ and 
 $n_{s } = 1-2/N_{e} \approx  0.967$ and shift it up by $0.006$, in accordance with  \cite{Akrami:2017cir}, we find
 \be
 n_s^{\rm quint} \approx  0.967+ 0.006 \approx 0.973 \ .
\label{quint} \ee
This estimate suggests that kination is able to make quintessential  $\a$-attractors fully compatible with the Planck+ACT +DESI  result $ n_s \approx 0.974 \pm 0.0034$ \cite{Louis:2025tst}.

 Over the last few years, additional studies of quintessential $\a$-attractors have been performed, stimulated by recent dark energy experiments and hints that dark energy might be evolving. The two-shoulder models which were used in new studies \cite{Akrami:2020zxw,Zhumabek:2023wka,Giare:2024sdl,Alestas:2024eic}
were proposed and studied in \cite{Akrami:2017cir}. They involve the linear exponential dependence of the potential on the geometric field $\phi=
 \sqrt {6 \alpha}\, \tanh{\varphi\over\sqrt {6 \alpha}}$.
 \be
 V_{{\rm Exp}-I}=M^2 e^{\gamma \big ({\phi\over \sqrt{6\a} }-1\big)}  \ .
\label{I} \ee
This model describes dark energy using an evolving equation of state $w$, which asymptotically approaches $w = -1$.
\be
 V_{{\rm Exp}-II}=M^2 e^{-2\gamma} \Big (e^{\gamma \big ({\phi\over \sqrt{6\a} }+1\big)}-1\Big) \ .
\label{II}  \ee
An amazing feature of the Exp-II model \rf{II} is that the asymptotic equation of state never reaches the value $-1$ and is given by
\be
w_{\infty} = -1 +{2\over 9 \a}  \ .
\ee

The results of the analysis in \cite{Akrami:2020zxw,Zhumabek:2023wka,Giare:2024sdl,Alestas:2024eic} appear to be consistent with some values of $3\a$ related to Poincar\'e disks and DES and DESI observations at the time when these studies were performed. For example, in \cite{Alestas:2024eic} it is suggested that the DESI data are compatible with the Exp-I model \rf{I} with $3\a=5$ or  $3\a=6$.

Here we only briefly discussed some basic single-field quintessential $\a$-attractors, where the same field is responsible for inflation and dark energy. However, two-field models of this type allow a much greater model-building flexibility and a wider choice of parameters \cite{Akrami:2017cir}.

\section{\boldmath Inflationary models with $n_{s} > 1-2/N_{e}$}

 Fig. \ref{Flauger0} and the closely related figures in the LiteBIRD paper \cite{LiteBIRD:2022cnt} show only three realistic targets for the future investigation: Starobinsky model, Higgs inflation, and various versions of $\alpha$-attractors.  These figures do not show any realistic targets with $n_{s} > 0.966$. As shown in the previous sections, the numerical bound $n_{s} > 0.966$ indicated in  Fig \ref{Flauger0} is not very strict because the attractor regime $n_{s} = 1-2/N_{e}$ is reached in the limit of small $\a$ and large  $N_{e}$.  The prediction $n_{s} = 1-2/N_{e}$, which is valid for the Starobinsky model, Higgs inflation, and $\alpha$-attractors, depends on $N_{e}$, which, in its turn, depends on the mechanism of reheating. If, for example, there is a long stage of kination after inflation, $n_{s}$ may become significantly higher than $0.966$ \cite{Akrami:2017cir}. A more general question, therefore, is whether there are any compelling targets for future observations with $n_s > 1-2/N_{e}$.

This issue was addressed in numerous publications following the Planck 2018 data release; see, e.g., \cite{Kallosh:2018zsi,Kallosh:2019hzo,Kallosh:2022feu,Kallosh:2022ggf} and references therein. For a while, this could seem to be a purely academic exercise since there was no demonstrated need to look for the models with $n_s > 1-2/N_{e}$. However,  as we already mentioned, the status of all models with $n_s = 1-2/N_{e}$ shown in Fig. \ref{Flauger0} was challenged by the latest ACT data release, which favored $n_s \approx 0.974$   \cite{Louis:2025tst,ACT:2025tim}.  Therefore, in this section, we will study several models with $n_s > 1-2/N_{e}$.

\subsection{Chaotic inflation with nonminimal coupling to gravity}\label{nonmchaot}

 We will begin with the basic chaotic inflation model $\tfrac12 {m^{2}\phi^{2}}$ with nonminimal coupling to gravity $(1+ \phi)R$ \cite{Kallosh:2025rni}, which was proposed immediately after the latest ACT data release. The Lagrangian of this model is
  \begin{equation}
{1\over \sqrt{-g}} \mathcal{L} =  \tfrac12 (1+\phi) R  - \tfrac12 (\partial \phi)^2 -  \tfrac12  m^{2}\phi^{2}  \ . \label{Jordan} 
 \end{equation}
This model belongs to a general class of $\xi$-attractors \rf{JordanGen} developed in our paper  \cite{Kallosh:2013tua} and discussed here in section \ref{sec:xi}.
A detailed investigation of inflationary predictions of this class of models has been performed in our paper \cite{Kallosh:2013tua} for  $V(\phi) = \phi^{2n}$ with various values of $n$ in a broad range of $\xi$, see Fig. \ref{xi} here. These models have a general prediction $n_{s} = 1-2/N_{e}$ in the large $\xi$ limit. As we already mentioned, for $n > 1$ this limit is reached very quickly, so that the attractor point $n_{s} = 1-2/N_{e}$ can be considered a general prediction of such models. However, for $n \leq 1$ this limit is reached relatively slowly. In this case, it makes sense to study predictions of $\xi$-attractors for various values of $\xi$, including the simplest case $\xi = 1$ as in \rf{Jordan}.

\begin{figure}
\centerline{\includegraphics[scale=.135]{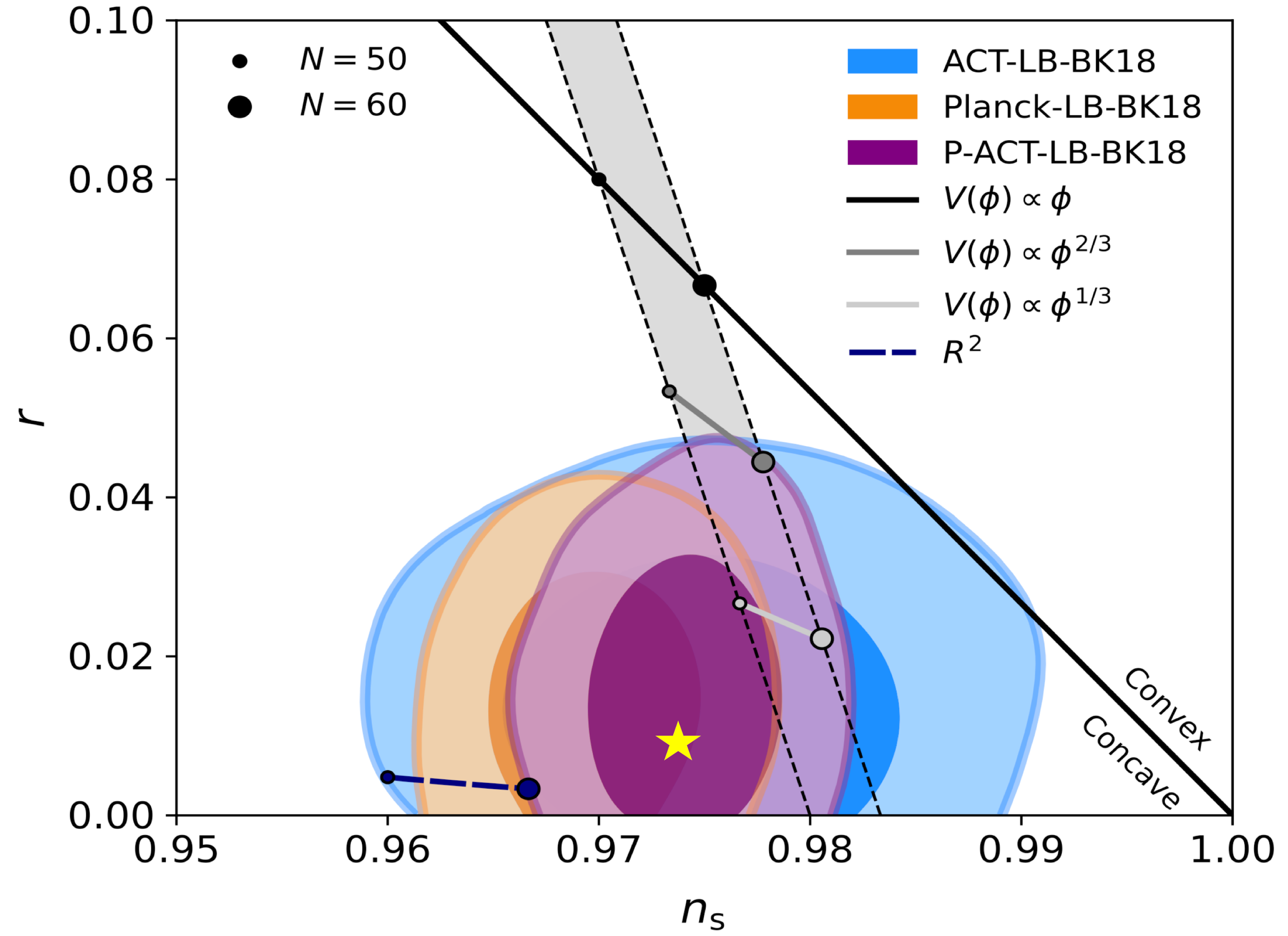}}
\caption{\footnotesize {The figure shows the latest constraints  on $n_{s}$ and $r$ according to ACT  (P-ACT-LB)  \cite{Louis:2025tst}. The dashed line at the bottom corresponds to the Starobinsky model. The yellow star at the center of the dark purple area favored by ACT shows the values of $n_{s}$ and $r$ \rf{60} in the  model  \rf{Jordan} for $N_{e}= 60$.}}
\vspace{-.3cm}
\label{xiStarAlone}
\end{figure}

An important feature of this scenario, which will appear repeatedly in several other models that we will describe, is that it has a plateau potential, but the plateau is approached not exponentially but polynomially: In the large $\phi$ limit, the potential is given by
\be
V = {m^{2}\over 2} \left(1 -{8\vp^{-2}}+O(\phi^{-4})\right) \ .
\ee
Using the results of  \cite{Kallosh:2013tua} in application to the simplest model \rf{Jordan}, one finds that 
\bea\label{60}
&&n_s= 0.9733, \quad r= 0.0094, \  \qquad  N_e=60 \ ,\\
&&n_s= 0.9679, \quad r= 0.0125,  \qquad \ N_e=50 \ .
\eea
The  $N_e=60$ predictions of this model are shown in Fig. \ref{xiStarAlone} by a yellow star close to the center of the dark purple area favored by ACT.
Tensor modes at this level ($r \sim 10^{-2}$) could be explored by BICEP/Keck \cite{BICEPKeck:2024stm} and by the Simons Observatory \cite{Hertig:2024adq}.
 
Extending these results to the model
 \begin{equation}
{1\over \sqrt{-g}} \mathcal{L}=  \tfrac12 (1+\xi \phi) R  - \tfrac12 (\partial \phi)^2 -  \tfrac12 {m^{2}}\phi^{2}  \ , \label{xi2} 
 \end{equation}
one finds the following approximate expression for $n_{s}$ and $r$ in the large $N_{e}$ approximation for $\xi = O(1)$:
  \begin{equation}\label{nsrxi}
n_s \approx 1- {3\over 2 N_{e}},   \qquad  r\approx {4\over \xi \, N_{e}^{3/2}}\ .
\end{equation}
This explains an increase of $n_{s}$ and $r$ as compared to the predictions of the Starobinsky model and Higgs inflation  $n_s= 1- {2\over   N_{e}}$,  $ r={12\over  N_{e}^{2}}$, and with respect to the predictions of $\alpha$-attractors $n_s= 1- {2\over   N_{e}}$,  $ r={12\alpha \over  N_{e}^{2}}$  \cite{Kallosh:2013yoa}.
Numerically,  the inflationary predictions of the model \rf{xi2} for  $N_{e} = 60$ are compatible with the P-ACT-LB constraints for $0.3< \xi <4$ \cite{Kallosh:2025rni}. 

As one can see from Fig.  \ref{xi}, predictions of other models with $n \lesssim 1$, including the model with $V\sim \phi$, are also compatible with the ACT constraints in a certain range of $\xi$;  see also a recent discussion of this issue in \cite{Gao:2025onc}.  

In the next subsection, we will discuss several other models that yield predictions compatible with the results of the latest ACT data release. They include 
pole inflation, KKLTI inflation, polynomial $\alpha$-attractors, and hybrid $\alpha$-attractors. These models are described in \cite{Kallosh:2019hzo,Kallosh:2022feu,Kallosh:2022ggf}. 

The main advantage of the model \rf{Jordan}  is its ultimate simplicity.  Just like the Starobinsky model and the model $\tfrac12  m^{2}\phi^{2}$, the model \rf{Jordan} has only one parameter, $m$, which determines the amplitude of the inflationary perturbations. 

On the other hand, its generalized version \rf{xi2} can also provide a good fit to the ACT data while allowing control of the tensor-to-scalar ratio $r$ by changing $\xi$. In this respect, the model \rf{xi2} is similar to $\alpha$-attractors, where one can control $r$ by changing the parameter $\alpha$.\footnote{In this context we should mention that some modified gravity models in the Palatini formalism are also compatible with ACT data, see e.g. \cite{Jarv:2017azx, Antoniadis:2018ywb,Barman:2023opy,Dioguardi:2025vci,Dioguardi:2025mpp}.
This class of models is very interesting but somewhat less explored. In particular, all models discussed in this paper up to now have supergravity generalizations, but we are unaware of any supergravity versions of inflationary models based on the Palatini formulation.}

\subsection{\boldmath Pole inflation, D-brane inflation, KKLTI models, polynomial $\alpha$-attractors}\label{sec:pole}

$\alpha$-attractors represent a special version of a more general class of attractors, known as pole inflation models \cite{Galante:2014ifa}. It is obtained by generalizing equation \rf{actionE}:
\be
{ {\cal L} \over \sqrt{-g}} =  {R\over 2} - {a_q\over 2} {(\partial \rho)^2 \over \rho^{q}} - V(\rho) \ .
\label{actionQ}\ee
Here the pole of order $q$ is at $\rho=0$ and the residue at the pole is $a_q$. For $q = 2$, $a_2 = {3\alpha \over 2}$, this equation describes E-models of $\alpha$-attractors, but here we   consider general values of $q$. For $q \not = 2$ one can always rescale $\rho $ to make $a_q=1$.  Just as in the theory of $\alpha$-attractors, one can make a transformation to the canonical variables $\varphi$ and find that the asymptotic behavior of the potential $V(\varphi)$ during inflation is determined only by  $V(0)$ and the first derivative ${dV(\rho)\over d\rho}|_{\rho = 0}$. The value of $n_{s}$ for this family of attractors  is given by 
\be
n_s= 1 - {\beta\over N_e}\, , \qquad \beta = {q \over q-1} \ .
\label{ns}
\ee
The models with  $\beta < 2$  provide  spectral index $n_{s}$ slightly greater than the $\alpha$-attractors result $n_s= 1 - {2\over N_e}$ . 

For all $\alpha$-attractors considered so far, the plateau of the potential is reached exponentially. For $q > 2$ the approach to the plateau is controlled by negative powers of $\varphi$.

 Some of these models described in  \cite{Kallosh:2018zsi} and \cite{Kallosh:2019eeu,Kallosh:2019hzo}   have interpretation in terms of Dp-brane inflation  \cite{Dvali:1998pa,Burgess:2001fx,Kachru:2003sx}. 
The ${\rm Dp}-{\overline {\rm Dp}}$  brane inflation plateau potentials are given by 
\be
V_{Dp-\bar Dp} \sim   {|\varphi|^k\over m^k + |\varphi|^k}= \Big( 1+  {m^k\over |\varphi|^k }\Big )^{-1}\, , \qquad 
m\ll 1\, , \qquad
k=7- p= {2\over 2-q} \ .
\label{KKLTI}
\ee
In {Encyclop\ae{}dia Inflationaris} \cite{Martin:2013tda}, these potentials were studied for arbitrary values of an unconstrained parameter $\mu$ replacing small $m$ of string theory. These were called KKLTI potentials.  

Their attractor formula for the $n_s$ is given in \rf{ns}, whereas the formula for $r$ depends on the parameter $m$ in the potential. 
For ${\rm D3}-{\overline {\rm D3}}$ inflation   for small $m$  one has   
\be\label{quart}
V = V_{0} \ {\varphi^{4}\over m^{4} + \varphi^{4}}\ , \qquad n_s =  1- {5 \over  3 N_{e}} \ , \qquad r = {4 m^{{4\over 3} }\over (3N_{e})^{5\over 3}} \ .
\ee
For  ${\rm D5}-{\overline {\rm D5}}$ brane inflation one has 
\be\label{D5}
V = V_{0} \ {\varphi^{2}\over m^{2} + \varphi^{2}}\ , \qquad n_s =  1- {3 \over  2 N_{e}} \ , \qquad   r=   {\sqrt 2\, m \over N_{e}^{3\over 2}} \ .
\ee
In explicit string theory constructions, the parameter $m$  is very small,  and $r$ is extremely small. For example, for the case $k=4$ in \cite{Kachru:2003sx} it was shown that 
\be
n_s\approx 0.97, \qquad r\approx 2\cdot 10^{-10} \ .
\ee
Most recently, brane-antibrane inflation models were revisited in \cite{Cicoli:2024bwq} and a new string realization was proposed, with
\be
n_s\approx 0.970, \qquad r\approx 2\cdot 10^{-8}  \ .
\ee
At small values of $m$, instead of the Brane Inflation  potentials \rf{KKLTI} one can use their small $m$ expansion  
\be \label{BI1}
 V_{BI} \sim V_{0}(1 -{m^{k}\over \vp^{k}}+... )  \ .
 \ee
 In {Encyclop\ae{}dia Inflationaris} \cite{Martin:2013tda}, the  potentials $V_{BI} \sim V_{0}(1 -{m^{k}\over \vp^{k}})$  were called BI potentials. 
 
 The numerical evaluation of observables KKLTI models is given in \cite{Martin:2013tda} for  $k=2, 3,4$ starting with $r< 10^{-1}$. In the subsequent work \cite{Kallosh:2018zsi}, the stringy origin of the inverse harmonic functions in eq. \rf{KKLTI}  was clarified, and a detailed comparison of BI   and KKLTI models of inflation was presented. It was emphasized that at small $r$, the predictions of the BI and KKLTI models coincide. However, at  $ r > 10^{-3}$ their values of $n_s$ may differ significantly. Moreover, it was realized that the values of $n_{s}$ in the KKLTI models have an attractor behavior: their values of $n_{s}$ do not depend on $r$ in the small $r$ limit. The embedding of these models into supergravity was also presented in \cite{Kallosh:2018zsi}, see also \cite{Pallis:2025gii}.
 
Some potentials that appear in the pole inflation scenario may have an alternative interpretation unrelated to Dp-branes.   For example, a quadratic model $V  \sim   {\varphi^2\over m^2 + \varphi^2}$ was first proposed in \cite{Stewart:1994pt}. It was used in \cite{Dong:2010in} as an example of a flattening mechanism for the $\varphi^2$ potential due to the inflaton interactions with heavy scalar fields. 
To explain the basic idea, consider a potential
\be
V(\phi, \chi) =  \phi^{2}\chi^{2} + m^2 (\chi-\chi_{0})^{2} \ .
\ee
If $m^{2}$ is sufficiently large, during inflation driven by the field $\phi$, the field $\chi$ will track its instantaneous minimum, depending on $\phi$, and the effective potential of the field $\phi$ along the inflationary trajectory becomes
\be
V = m^{2} \chi_{0}^{2}\  { \phi^{2}\over   \phi^{2} +m^{2}}\  .
\ee
All potentials \rf{KKLTI} discussed above can also be obtained in the context of {\bf polynomial $\alpha$-attractors} \cite{Kallosh:2022feu} where
\be
{ {\cal L} \over \sqrt{-g}} =  {R\over 2} -{3\a\over 2} {(\partial \rho)^2 \over \rho^{2}} - V_0\,   {\ln^k\rho \over  \ln^k\rho+ 1 } \ .
\ee
By the change of variables  $\rho = e^{-\sqrt {2 \over 3\alpha}\varphi}$, as in  E-models \rf{apole}, \rf{Emodel},  this model is reduced to a model for a canonical scalar field with the potential 
\be
V_{\rm poly}(\vp) =V_0 {|\varphi|^k\over \mu^k + |\varphi|^k}\, , \qquad \mu = \sqrt{3\alpha\,\over 2}  \ .
\label{polynomial}\ee
 Thus, polynomial  $\alpha$-attractor models  at small values of $\mu$ are the same as  ${\rm Dp}-{\overline {\rm Dp}}$  BI models.  

A more general form of the potential which can be used for polynomial $\a$-attractors for odd or non-integer  $k$  is $V_{0}\, {(\ln^2\rho+1)^{k/2} -1 \over  (\ln^2 \rho+1)^{k/2} +1 } $. In the large $\vp$ limit, this allows us to reproduce the potential   \rf{polynomial} for arbitrary positive values of $k$.
\begin{figure}[H]
\centering
\includegraphics[scale=0.3]{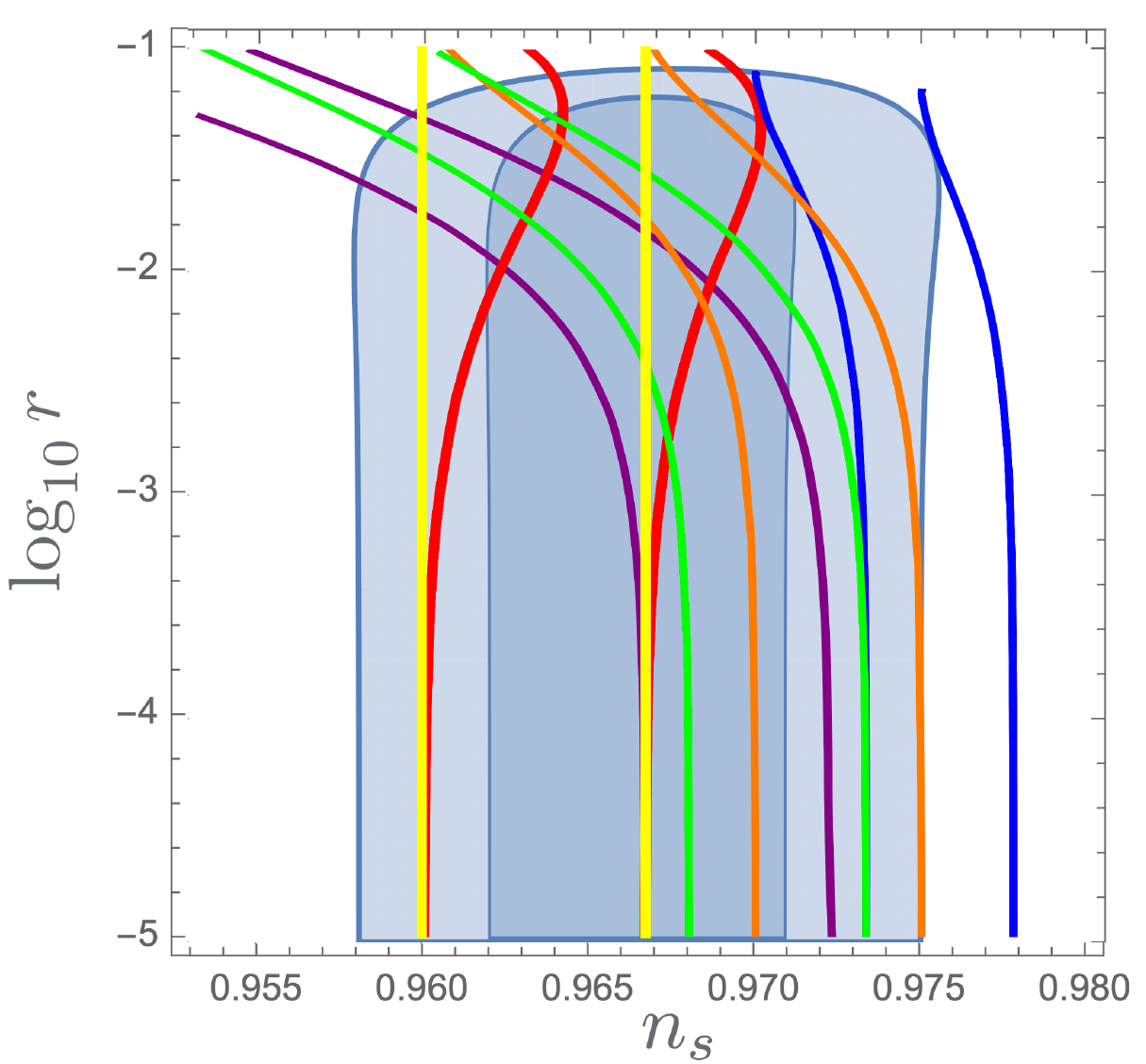}
\vskip -5pt
\caption{\footnotesize A combined plot of the predictions of the simplest exponential $\alpha$-attractor models and polynomial $\alpha$-attractor models (KKLTI models)  for $N_{e} = 50$ and $60$ \cite{Kallosh:2019hzo}. From left to right, we show predictions of T-models and E-models (yellow and red lines) and of the KKLTI models  (purple, green, orange, and blue lines) for potentials in eq. \rf{KKLTI} with $k=4,3,2,1$. The blue data background corresponds to the Planck 2018 results, including BAO. Below $r\approx 10^{-3}$ the polynomial $\alpha$-attractor models have the same predictions as $Dp-\overline {Dp} $ models with $p=3,4,5,6$. Thus, exponential \cite{Kallosh:2013yoa} plus polynomial $\a$-attractors  \cite{Kallosh:2022feu} cover the total blue area in this plot.}
\label{Blue}
\end{figure}
At large $\vp$, the potential in \rf{polynomial}  and its predictions for $n_{s}$ are
\be \label{pol2}
 V \sim V_{0}(1 -{\mu^{k}\over \vp^{k}}+... ) \ , \qquad n_{s} = 1-{2\over N_{e}}{k+1\over k+2} \ .
 \ee
Note that for any $k> 0$ the attractor values of $n_{s}$ in \rf{pol2} are greater than $1-2/N_{e}$, and in the limit $k \to 0$ one has $n_{s} \to  1-1/N_{e}$.
This means, for example, that for $N_{e} = 55$, this class of models can account for $n_{s}$ in the broad range $0.9636< n_{s} < 0.982$.
In Fig. \ref{Blue}, we give a combined plot of the predictions of the simplest exponential $\alpha$-attractor models and polynomial $\alpha$-attractor   inflationary models for $n_{s}$ and $\log_{10}r$,  for $N_{e} = 50$ and $60$ \cite{Kallosh:2019hzo}.  In the small $\mu$ limit, the predicted values of $r$ for all these models can take extremely small values, all the way down to $r \to 0$. 

Independently of their interpretation, exponential and polynomial $\alpha$-attractors  may serve as a powerful tool for parametrization of all observational data with $ 1-2/N_{e} \lesssim n_{s}< 1-1/N_{e}$,  i.e. in the range $0.96 < n_{s}<  0.98$ for $N_{e} = 50$, or   $0.967 \lesssim n_{s}  \lesssim 0.983$  for  $N_{e} = 60$ \cite{Kallosh:2019eeu,Kallosh:2019hzo}. 

\begin{figure}[H]
\centering
\includegraphics[scale=0.2]{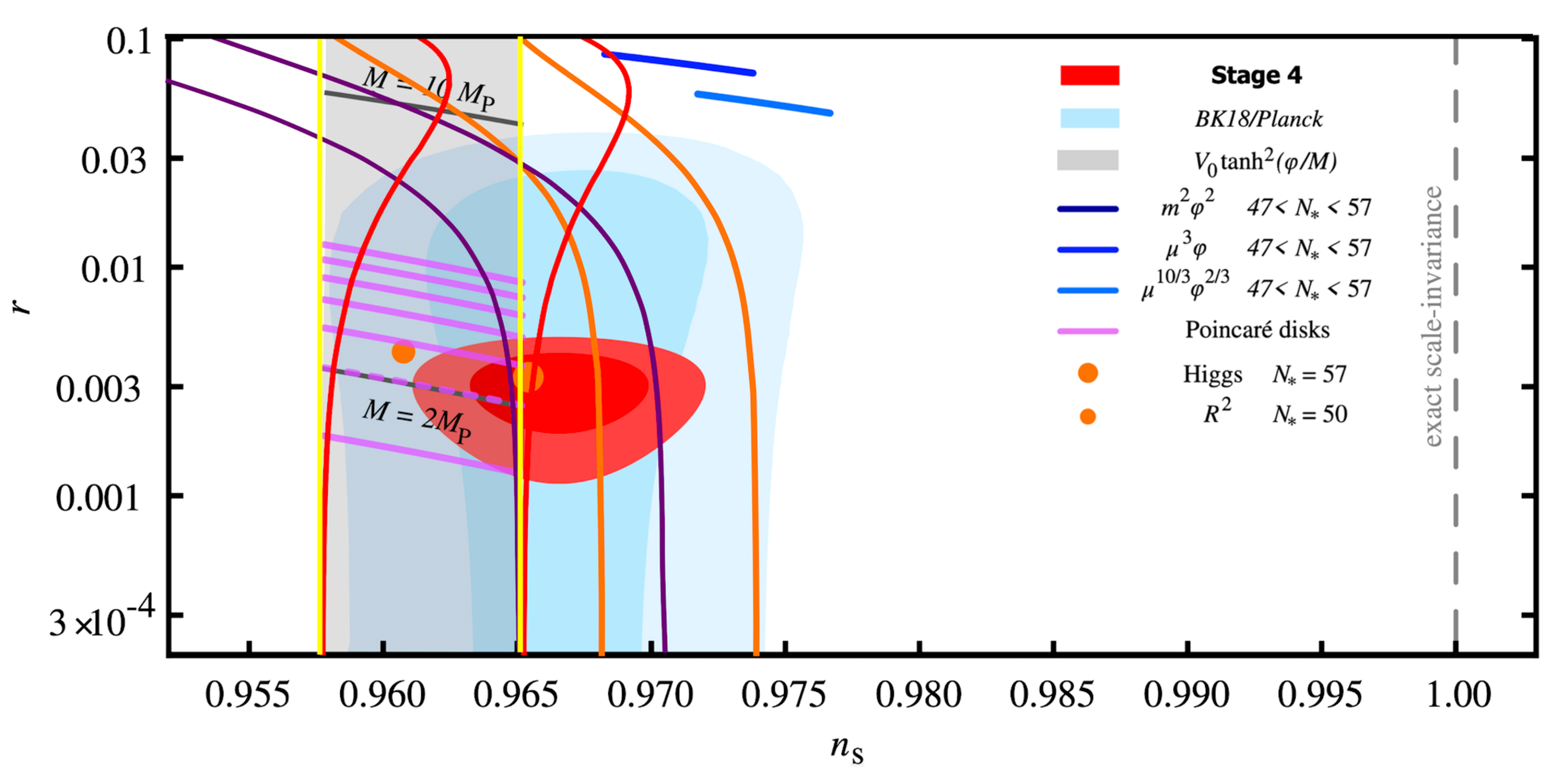}
\caption{\footnotesize An extended version of the figure from  ``Snowmass2021 Cosmic Frontier: Cosmic Microwave Background Measurements White Paper'' \cite{Chang:2022tzj} which was presented in  Fig. 1 in \cite{Kallosh:2022vha}. It shows the predictions of T-model $\alpha$-attractors with unconstrained values of $\alpha$ (gray area), E-models (red lines), the predictions for $3\alpha= 7,6,5,4,3,2,1$ (purple lines), as well as Higgs inflation, $R^2$ inflation (red dots). In addition, the purple lines show the prediction of the polynomial attractors \rf{quart}, and the orange lines correspond to \rf{D5}. The predictions are for $47 < N_e< 57$.}
\label{Flauger}
\end{figure}

As illustrated by Fig. \ref{Flauger}, just a few of such stripes may completely cover a broad range of possible values of $n_{s}$ and $r$ compatible with the Planck2018  data. The left part of this range is well covered by exponential  $\alpha$-attractors. The right part of this range is well covered by the models discussed in this section. This works especially well in the small $r$ limit, which is the top priority for parametrizing the results of the ongoing and planned search for the inflationary gravitational waves.

Finally, we should mention a distinguishing feature of the polynomial $\alpha$-attractors. Even though these models describe the potentials of the same type \rf{polynomial} as all other models described in this section, the polynomial $\alpha$-attractor have the same geometric origin as all other $\alpha$-attractor models, and therefore they also have a discrete series of preferred values $3\alpha = 1, 2, ..., 7$. As a result, for each value of $k$, we have a series of predictions for $r$ corresponding to these values of $3\alpha$. For example, in the simplest model with $k = 2$ we have the following Poincar\'e disk series:
\be\label{poinpol}
n_s =  1- {3 \over  2 N_{e}}, \qquad  r = {\sqrt{3\alpha}\over N_{e}^{3/2}}, \quad 3\alpha = 1, 2, ..., 7 \ .
\ee 
In particular, for $N_{e} = 60$ we have 7 specific predictions for $r$:
\be\label{poinpoln}
n_s =  0.975, \qquad  r = 0.0021,\, 0.0030,\,  0.0037,\,  0.0043, \, 0.0048, \, 0.0052, \, 0.0056\ .
\ee

\section{Hybrid attractors}

The original hybrid inflation model  \cite{Linde:1991km,Linde:1993cn} has the potential
\begin{equation}\label{hybrid}
V(\sigma,\phi) =  {1\over 4\lambda}(M^2-\lambda\sigma^2)^2
+ {m^2\over
2}\phi^2 + {g^2\over 2}\phi^2\sigma^2\ .
\end{equation}
 The effective mass squared of the field $\sigma$ at $\sigma = 0$ is $V_{ \sigma,\sigma}(\sigma = 0) = -M^{2}+g^{2}\phi^{2}$.
For $\phi > \phi_c = M/g$ the only minimum of the effective potential
$V(\sigma,\phi)$ with respect to $\sigma$  is $\sigma = 0$. The curvature of the effective potential in the $\sigma$-direction is chosen to be much greater than in the  $\phi$-direction. At the first stages of expansion of the Universe, the field $\sigma$ rolls down to $\sigma =
0$, whereas the field $\phi$ remains large and drives inflation.

The potential at $\sigma = 0$ is given by
\be
V(\sigma= 0,\phi) = {M^4\over 4\lambda} +  {m^2\over 2}\phi^2 \ .
\ee
It is the simplest chaotic inflation potential $ {m^2\over 2}\phi^2$ uplifted by  
$V_{\rm up} = {M^4\over 4\lambda} $. If uplift is sufficiently large, inflation may continue all the way down to $\phi = 0$. However, when the inflaton field $\phi$ rolls down and becomes smaller than  $\phi_c = M/g$, the effective mass squared of the field $\sigma$ becomes negative, and the phase transition with the spontaneous symmetry breaking occurs. For a proper choice of parameters, this phase transition occurs very fast, and inflation abruptly ends \cite{Linde:1991km,Linde:1993cn}. 

In the original version of this model, the uplift is large, $n_{s}$ is typically very close to 1, and $r$ can be extremely small. Indeed, expressions for $1-n_{s}$ and $r$ contain terms proportional to $(V'/V)^{2}$ and $V''/V$. For any given $\phi$, the uplift $V_{\rm up}$ does not change the derivatives of the potential while increasing $V$, which decreases $1-n_{s}$ and $r$.  But if the uplift is very small, inflation occurs as in the simple chaotic inflation model  ${m^2\over 2}\phi^2$, and $n_{s} = 1-2/N_{e}$.

A similar situation appears if one embeds the model \rf{hybrid} in the context of $\alpha$-attractors \cite{Kallosh:2022ggf}. In this class of models, one has two basic attractor regimes. If $V_{\rm up}$ is large and $\phi_{c} \ll  1$, then the last stage of inflation occurs as in the original hybrid inflation scenario,  with $n_{s}$  very close to 1. On the other hand, if uplifting is small, one has the standard $\alpha$-attractor result  $n_{s} = 1-{2\over N_{e}},  r = {12\over N^{2}_{e}}$.  Thus, by changing the parameters of the model, one may interpolate between $n_{s} = 1-{2\over N_{e}}$ and the  $n_{s} = 1$ \cite{Kallosh:2022ggf,Iacconi:2024hmg}.

The full theory of this effect is rather nuanced as it depends on relations between many parameters of the $\alpha$-attractor version of the model \rf{hybrid}. For example, in certain cases, the process of spontaneous symmetry breaking can drive a short second stage of inflation, resulting in the production of large primordial black holes \cite{Braglia:2022phb}, which might even account for dark matter \cite{Garcia-Bellido:1996mdl}. For some other relations between the parameters, the minimum of the potential at $\sigma = M/\sqrt \lambda$ can never be reached, and the standard Higgs-type symmetry-breaking potential of the field $\sigma$ transforms into a quintessential inflation potential shown in the right part of  Fig. 4 in \cite{Kallosh:2022ggf}. 

The general lesson is that a theory of multifield $\alpha$-attractors may have more than one attractor regime, depending on the choice of the parameters. This takes away some of the predictive power of single-field $\alpha$-attractors, but brings many new interesting opportunities, including the possibility to have $n_{s}$ in the range matching the ACT results, or even much higher.

\section{String theory inflation}\label{sect:string}

\begin{figure}[H]
\centering
\includegraphics[scale=0.7]{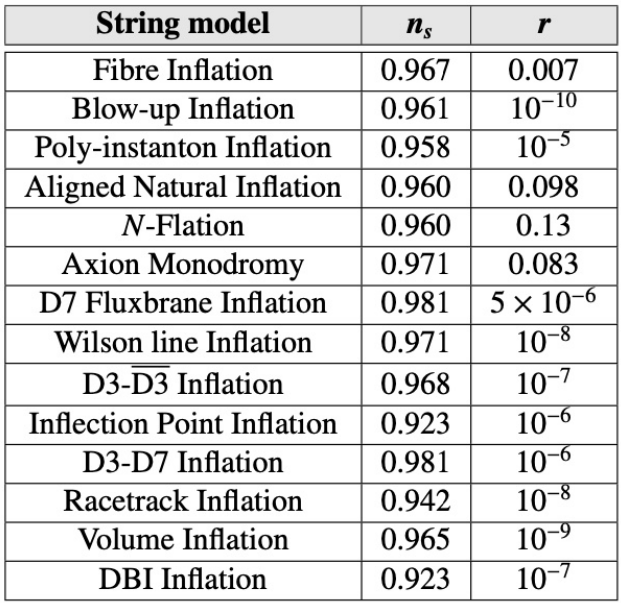}
\caption{\footnotesize Predictions of string inflation models according to the review of string inflation  \cite{Cicoli:2023opf}.}
\label{Table}
\end{figure}

String theory inflation is an important topic that was reviewed in detail two years ago in  \cite{Cicoli:2023opf}.  At that time, the main models of string inflation, evaluated at $N_e \simeq 52$, were given in the table presented here in Fig. \ref{Table}.  
Here, we will make a few comments regarding the status of recent string theory inflation models and observational results.

First of all, the Aligned Natural inflation, N-flation model, Inflection Point model, the Racetrack inflation, the DBI model, and the version of the Axion Monodromy model given in the table are practically ruled out by the data. That is 6 models out of the 14 models mentioned in the table.

Most of the remaining string inflation models in Fig. \ref{Table} have tensor-to-scalar ratio $r\leq 10^{-5}$, far beyond current observational reach. For example,  the goal of  LiteBIRD is to reach $r\sim 2\times 10^{-3}$. 
Nevertheless, all models can be tested by their predictions for $n_s$, which will require both theoretical and experimental clarification in the future.  

Here, we will mention two recently updated models and one model that is not present in Fig. \ref{Table}.

1. Fibre inflation, see  \cite{Cicoli:2024bxw} and references therein, has the effective potential  
\be
V=V_{0} (1-c_1\ e^{-{1\over \sqrt 3}\vp}) \ ,
\label{exp}\ee
which is the  $\a$-attractor potential with $\a=2$ \cite{Kallosh:2017wku}. An updated  version of this model, discussed in  \cite{Cicoli:2024bxw},  has the following predictions
\be\label{fbr}
 n_s= 0.963, \qquad r\approx 0.0072, \qquad  N_e = 50.
\ee
Even if confirmed by future observations, this model will not unambiguously support string theory ``from the sky'', being degenerate with $\a$-attractors for $\alpha = 2$.  Moreover, in both cases, there is an issue raised by the ACT constraint $n_{s} =0.9743 \pm  0.0034$. Future data on $n_s$ from SPT-3G and the Simons Observatory, as well as theoretical developments, will be crucial in this respect.

2.  The updated string theory inflation model known as the Loop Blow-up Inflation \cite{Bansal:2024uzr}
\be
V=V_0 (1-c_2\, \vp^{-2/3}) \ .
\label{pol}\ee 
 The predictions of the specific string theory realization of this model  \cite{Bansal:2024uzr} are
\be 0.975\leq n_s\leq 0.976, \qquad r\approx 2\times 10^{-5}, \qquad 51.5\leq N_e \leq 53   \ .
\ee
Whereas the value of $n_s$ is compatible with the ACT constraint $n_{s} =0.9743 \pm  0.0034$, the value of $r$ is practically unobservable.

3. The  axion monodromy model \cite{McAllister:2014mpa} with the potential $\vp^{2/3}$. (The table describes a different version of the axion monodromy model with a linear potential.) The version with the potential $\vp^{2/3}$ predicts, for $N_{e} = 60$, 
\be
n_s\approx 0.978\, ,\qquad r\approx 0.044 \ .
\ee
The Planck/BICEP/Keck data shown in Fig.  \ref{BICEPKeck} strongly disfavored it, but in view of the recent ACT data, the status of this model somewhat improved.   It is at the boundary of the $2\sigma$ (light purple)  P-ACT-LB-BK18 region shown in Fig. \ref{ACT}, so this model is marginally consistent with ACT.

\vskip 15pt

\section{Discussion}

In this short review, we have mentioned many different inflationary models, and yet there are many others which can be found in the {Encyclop\ae{}dia Inflationaris} \cite{Martin:2013tda}. It is difficult to make any definite conclusions based on our subjective choice of the models to discuss, and it is especially difficult in a situation where there is an ongoing discussion of the $H_{0}$ problem and of the validity of the basic $\Lambda$CDM model. 

However, we will do our best and update this article if new data forces us to reconsider our conclusions. We will divide this Discussion into several different parts.

\vskip 7pt

{\bf   Do we have a simple, comprehensive inflationary model that works no matter what?}

The answer is Yes and No. If our only goal is to find a model describing any values of the 3 CMB-related parameters $A_{s}$, $n_{s}$ and $r$, including the Planck/BICEP/Keck results and the recent ACT results, this model is already available and it was known for nearly 2 decades: One can always find a polynomial potential with 3 parameters which fits any values of $A_{s}$, $n_{s}$ and $r$ \cite{Destri:2007pv,Nakayama:2013jka,Kallosh:2014xwa}. Moreover, one can easily do it in supergravity  \cite{Kallosh:2014xwa}, see Section~\ref{sec:polyn} in this paper.

However, one of our goals pursued in this paper was to describe simple but well-motivated models that could describe all CMB-related data by a choice of a single parameter. Also, the polynomial model mentioned above is a single-field model, as are most other models discussed in this paper. A discovery of local non-Gaussianity $f_{NL} > O(1)$ would force us to consider more complicated models. This highlights the importance of investigating non-Gaussianity; see, e.g., \cite{Heinrich:2023qaa, Shiveshwarkar:2023afl}.  

 \vskip 7pt

{\bf Can we test string theory inflation?}

The string inflation models have already been tested. As we mentioned in Sec. \ref{sect:string},  6 models out of the 14 models presented in the table \ref{Table} are already ruled out. Regarding the remaining models, two distinct possibilities emerge if inflationary tensor modes are discovered.

a) Suppose we find that $n_s= 0.963$ and discover tensor modes with $ r\approx 0.0072$, as  in  \rf{fbr}. This would be a great triumph of inflationary theory and of Fibre inflation, and/or of the $\alpha$-attractors with $\alpha = 2$.

b) Suppose instead that observations find tensor modes with any value of $r$ in the very broad range  $10^{{-4}} < r <0.036$,  but far from the point  $n_s= 0.963$, $r\approx 0.0072$ favored by Fibre inflation. Rather paradoxically, this discovery would rule out {\bf all} string inflation models mentioned in the review article    \cite{Cicoli:2023opf}.

This does not mean that it is impossible to construct a string theory model consistent with the ACT constraint and predicting $r$ in the range $10^{{-4}} < r < 0.036$. An example of such a single-inflaton model was proposed recently in \cite{Leontaris:2025hly}.  This model relies on  a judicious choice of more than ten different parameters.

\vskip 7pt

{\bf Can we implement large field inflation in supergravity?}

For a long time, there was a lore that one cannot embed large field inflation in supergravity because of the ``inevitable'' factor $e^{K}$ in the potential, with the Kahler potential $K = \Phi\bar\Phi$. This problem was solved in 2000 in \cite{Kawasaki:2000yn} for the simplest chaotic potential $m^{2}\phi^{2}$, and then the solution was extended to a broad class of chaotic inflation potentials in \cite{Kallosh:2010xz}. Subsequent developments using the nilpotent superfield resulted in constructing consistent models of inflation in supergravity with arbitrary inflationary potentials, including the large-field plateau potentials describing the Starobinsky model, Higgs inflation, and $\alpha$-attractors, see a recent review \cite{Kallosh:2025jsb}.

\vskip 7pt

{\bf Planck, ACT, and exponential vs polynomial approach to a plateau}

Inflationary models favored by Planck 2018, such as the Starobinsky model, Higgs inflation, and $\alpha$-attractors, have one common feature. Their potentials have a plateau which is approached exponentially. In particular, for $\alpha$-attractors
\be\label{plateau}
V_{\rm exp} (\vp) = V_{0} \left(1-  e^{-\sqrt{2\over 3\alpha} \varphi } +\dots  \right)   .
\ee
There may be some model-dependent pre-exponential factor, but it can be absorbed into a redefinition (shift) of the field $\vp$, see e.g. \rf{plateau1}. That is why the predictions of $\alpha$-attractors depend only on $V_{0}$ and $\alpha$, and are very stable with respect to even significant modifications of the original potential $V(\phi)$ and quantum corrections \cite{Kallosh:2016gqp}.

Note that the value of $n_{s}$ for all models with the potential approaching the plateau exponentially does not depend on the factor $\sqrt{2\over 3\alpha}$ in the exponent. That is why so many different models have the same universal prediction $n_{s} = 1-2/N_{e}$. For $N_{e}= 60$ this prediction gives $n_{s} \approx 0.967$, and for $N_{e} = 57$, as in the CMB-S4 figure  Fig. \ref{Flauger0}, $n_{s} \approx 0.965$.

Many models attempting to match the recent ACT constraint $n_{s} = 0.9743 \pm 0.0034$ have plateau potentials where the potential approaches the plateau not exponentially, but rather polynomially. For example, for polynomial $\alpha$-attractors \cite{Kallosh:2022feu} 
\be \label{p}
V_{\rm pol} (\vp)= V_{0} \left(1 -  \Big(\sqrt{2\over 3\alpha} \varphi\Big )^{-k} +\dots \right)   .
 \ee
The difference between exponential and polynomial models is shown in Figs. \ref{Blue}, \ref{Flauger}. At $r\gtrsim 10^{-2}$, the values of $n_{s}$ in many of the models are still far from the attractor points. But at about $r\lesssim 10^{-3}$, most of the models reach the attractor regime: $n_{s}$ does not change with the further decrease of $r$.  All exponential $\alpha$-attractors are on the left side of the blue area favored by Planck2018, whereas all polynomial $\alpha$-attractors are on the right side of this area.  They, as well as other models discussed in section \ref{sec:pole}, can describe models with  $n_{s}$   from 
 $1-2/N_{e}$ to $1-1/N_{e}$, i.e. in the range $0.967 \lesssim n_{s} \lesssim 0.983$ for $N_{e} = 60$.
 
By increasing the precision of determination of $n_{s}$, one may be able to distinguish the two different classes of inflationary models with qualitatively different families of potentials \rf{plateau} and \rf{p}.
Note that Simons Observatory plans to halve the current error bar on the scalar perturbation spectral index $n_{s}$ \cite{SimonsObservatory:2025wwn}.

Precise determination of $n_{s}$ could be sufficient for distinguishing between different classes of models even if they predict too small values of $r$ to be ever observed, because the predictions of $n_{s}$ become increasingly accurate in the small $r$ limit, see Fig. \ref{Blue}. This may help us to distinguish not only between the general classes of models  \rf{plateau} and \rf{p}, but also between different ${\rm Dp}-{\overline {\rm Dp}}$  brane inflation plateau potentials \rf{KKLTI} and the  Loop Blow-up Inflation \rf{pol}.

Moreover, knowledge of $n_s$   may help us to fine-tune the search for the tensor modes.  For example, for $n_{s} \lesssim 0.967$ we have 7 discrete targets for the future search shown in Fig. \ref{Flauger0}, in the range from $10^{{-2}}$ to $10^{{-3}}$.  Meanwhile, for $ n_s \gtrsim 0.97$, we also have a similar series of 7 different predictions for each value of the parameter $k$ in the theory of polynomial attractors, see \rf{poinpol}, \rf{poinpoln} for $k = 2$. Pole inflation models allow for a broad range of $n_{s}$ and $r$. On the other hand, all string inflation models with $r < 0.36$ and  $n_s \gtrsim 0.97$, as well as all ${\rm Dp}-{\overline {\rm Dp}}$ models,  yield tensor-to-scalar ratio in the unobservable range $r \lesssim 10^{{-5}}$.

\vskip 7pt

{\bf Did we miss anything? Yes....}

As we have already mentioned, there are many other models that deserve discussion in this review; see {Encyclop\ae{}dia Inflationaris} for a much longer and yet unavoidably incomplete list \cite{Martin:2013tda}. Each of the models is there for a good reason. Moreover, the discussion so far has only briefly mentioned potential seismic shifts that are still possible in cosmology. The examples are abundant: the $H_{0}$ problem, dark energy models challenging $\Lambda$CDM, concerns about possible anisotropies of the cosmological background, etc. In this ever-changing situation, it is encouraging that some inflationary models developed long ago still remain consistent with observations.

For example, a few years ago, it was difficult to expect that the theory of $\alpha$-attractors could simultaneously describe inflation and dark energy, and increase $n_{s}$ \cite{Dimopoulos:2017zvq,Akrami:2017cir}. Basic models of hybrid inflation \cite{Linde:1991km,Linde:1993cn} were ruled out long ago because they predicted $n_{s}$ too close to $1$, but now, in combination with cosmological attractors, hybrid inflation models may describe a broad range of $n_{s}$ compatible with Planck as well as with ACT \cite{Kallosh:2022ggf},  and they may even be responsible for production of primordial black holes and stochastic gravitational wave background  \cite{Braglia:2022phb}. Among the latest surprises is the intriguing possibility that curvatons \cite{Linde:1996gt,Enqvist:2001zp,Lyth:2001nq,Moroi:2001ct,Bartolo:2002vf,Linde:2005yw,Byrnes:2016xlk} can make the basic chaotic inflation scenario compatible with ACT data  \cite{Byrnes:2025kit}. Another possibility to achieve the same goal is to add the term $\phi R$ to the simplest chaotic inflation scenario $ \tfrac12 m^{2}\phi^{2}$ \cite{Kallosh:2025rni}.

We live in interesting times,  but being optimists (most of the time anyway), we consider it not a curse but a blessing.

\section*{Acknowledgement}
We acknowledge stimulating discussions with R. Bond, C. Burness, J.J. Carrasco, M.Cort\^es, G.~Efstathiou, F. Finelli, W. Freedman, C. Hill, C.L. Kuo, A. Liddle, V. Mukhanov, F. Quevedo, D. Roest,  E. Silverstein, T.~Wrase, and Y. Yamada.   This work was supported by the Leinweber Institute for Theoretical Physics at Stanford and by NSF Grant PHY-2310429.  It is an extended version of the talk given at the Lema\^itre Conference 2024 ``Black Holes, Gravitational Waves and Space-Time Singularities'' at Specola Vaticana. We are grateful to the organizers for their support and hospitality.
 
\appendix

\parskip 5pt

\section{General multi-field inflationary  equations for $\phi^i(N)$}\label{App:A}

 Most of inflationary models we discuss in this paper  are single field inflationary models. However, they often originate from multi-field models with curved space geometry. We have found it useful to describe background inflationary trajectories for such models either as evolution in time, $\phi^i(t)$ or as evolution  in  $N$,   $\phi^i(N)$.
 Here the number of e-foldings is defined as $N\equiv \ln{a(t_{\rm end})\over a(t)}$.

 The standard inflationary background equations in theories with an arbitrary curved scalar field geometry and a space-time FRW metric are well known.  An advantage of using these  equations with evolution in $N$ shows up, in particular,  in numerical studies of single field models where potentials are simple for non-canonical fields. Examples in this paper 
 where we used this setup are in Secs. \ref{sec:xi}, \ref{nonmchaot}.

Consider  cosmological models  in FRW space-time where the field space  has a general type curved geometry ${g}_{ij}(\phi )$
\begin{equation}
 \sqrt{-g}^{-1}   \mathcal{L}=  \left[-\frac{1}{2} R+ \frac{1}{2} {g}_{ij}(\phi ) \,\partial_\mu \phi^i \partial^\mu \phi^j -V(\phi)\right] \;.
\label{S}\end{equation}
Exact background equations of motion  for the homogeneous fields $\phi^i(t)$   are 
\begin{equation}\label{{eomt}}
\ddot\phi^i+3H\dot\phi+\Gamma^i_{jk} \dot \phi^j \dot \phi^k  +g^{ij}{\partial V\over \partial \phi^j}=0 \ ,
\end{equation}
where
\be
H(t) \equiv  {\dot a\over a} = \sqrt {V+ E_{\rm kin}\over 3}\, , \qquad E_{\rm kin}=  \frac{1}{2} {g}_{ij}(\phi ) \,\partial_\mu \phi^i \partial^\mu \phi^j \, .
\ee
These equations were  presented  in 
\cite{Kallosh:2002gf} (we skip here the cold dark
matter energy density).

The standard form of inflationary slow-roll parameters  can be given in the form \cite{Lyth:2009imm,Iacconi:2021ltm}  
\be
\epsilon_H=  - {\dot H\over H^2}\, , \qquad \eta_H= \epsilon_H-{1\over 2} { \dot \epsilon_H\over  H \epsilon_H} \ .
\ee
For single-field models, the observables are
\be
n_s= 1- 4\epsilon_H + 2\eta_H \ , \qquad r= 16 \epsilon_H \ . 
\label{nsN}\ee
To switch from time evolution to the number of e-foldings $N$ evolution, we use a relation
\be
dN\equiv d\ln a= H dt \ ,
\label{dN}\ee
We use $'$ for  ${\partial\over \partial N}$ derivatives, instead of dot for ${\partial\over \partial t}$.
The inflationary slow-roll parameters are 
\be
\epsilon_H= -{ H'\over H}\, , \qquad \eta_H= \epsilon_H -{1\over 2} { \epsilon'_H\over  \epsilon_H} \ ,
\ee
and observables depend on the number of e-foldings $N$.
After a change of variables according to Eq. \rf{dN} our inflationary background equations with $N$ as an evolution parameter take the form
\begin{equation}\label{eomN}
\phi''{}^{i}+(3-\epsilon_H)\phi'^{i}+\Gamma^i_{jk}  \phi'^{j}   \phi'^{k}   +{g^{ij}\over H^2} {\partial V\over \partial \phi^j}=0 \ .
\end{equation}
By solving it one can find a solution $\phi^i(N)$.
These equations are supplemented by the following relations
\be
H^2= { V\over 3-\epsilon_H}\, , \qquad 
\epsilon_H= -{ H'\over H}= \frac{1}{2} {g}_{ij}(\phi ) \, \phi'^{i}\phi'^{j} \ .
\ee
Analogous equations were presented in \cite{Peterson:2010np} where it was also shown that in the case of the two-field models, the extraction of observables like $n_s, r$ is complicated and depends on the turn rate of the 2-field background trajectory.
\

{\it Simple examples in single field case:}

\noindent  For the case of models with a nonminimal coupling to gravity \cite{Kallosh:2013tua} described here in Sec. \ref{sec:xi}, one can use eq. \rf{Einstein}, where the single scalar   $\phi$ in the Einstein frame is not canonical,  but the potential is given by a well-defined function of the original field $\phi$. Namely, the kinetic term of the scalars involves a geometry $g_{11}d\phi^1 d\phi^1$ 
\be
V_E= {V_J\over \Omega^2} \ , \qquad g_{11} = \Omega^{-1} +{3\over 2} (\log \Omega)^{' 2} \ .
\ee
Thus, we have a  single evolution equation where $\phi$ depends on $N$:
\begin{equation}\label{eomN2}
\phi''+(3-\epsilon_H)\phi'+\Gamma^1_{11}  \phi'   \phi'   +{g^{11}\over H^2} {\partial V\over \partial \phi}=0 \, ,   \qquad \Gamma^1_{11}= {1\over 2} \partial_\phi (\ln(g_{11})) \ .
\end{equation}
This case was studied for models with $\Omega= 1+\xi \phi^{n/2}$ in \cite{Kallosh:2013tua}. It is very simple to calculate $n_s$ and $r$ using the method described above. This procedure was also used for the model of the chaotic inflation with nonminimal coupling to gravity in Sec. \rf{Jordan}. 
\bibliographystyle{JHEP}
\bibliography{lindekalloshrefs}
\end{document}